\documentclass[a4paper]{article}
\usepackage{jheppub}
\usepackage{import}

\usepackage{hyperref}
\usepackage{mathtools}
\usepackage{natbib}
\usepackage{physics}
\usepackage{tikz}
\usepackage{tikz-cd}

\allowdisplaybreaks

\let\originalleft\left
\let\originalright\right
\renewcommand{\left}{\mathopen{}\mathclose\bgroup\originalleft}
\renewcommand{\right}{\aftergroup\egroup\originalright}

\newcommand{\eff}{\mathrm{eff}}
\newcommand{\JK}[1]{\underset{\small x=#1}{\operatorname{JK-Res}}}
\newcommand{\pic}[2]{\mathrm{Pic}^{#1}\left(#2\right)}

\DeclareMathOperator{\sign}{sign}
\DeclareMathOperator{\ch}{ch}
\DeclareMathOperator{\td}{td}
\DeclareMathOperator{\vol}{vol}

\usepackage{amsfonts}
\usepackage{amscd}
\usepackage{amssymb}
\usepackage{amsmath,bbm}
\usepackage{graphicx}
\usepackage{epsfig}
\usepackage{latexsym}
\usepackage{mathtools}
\usepackage{hyperref}
\usepackage{tikz-cd}
\usepackage[vcentermath]{youngtab}
    
\def \be  {\begin{equation}}
\def \ee  {\end{equation}}
\def \bea {\begin{equation}\begin{aligned}}
\def \eea {\end{aligned}\end{equation}}
\def \ba  {\begin{eqnarray}}
\def \ea  {\end{eqnarray}}
\def \bb  {\begin{thebibliography}}
\def \eb  {\end{thebibliography}}
\def \lab #1 {\label{#1}}

\newcommand\ep{\epsilon}

\newcommand\cC{\mathcal{C}}

\newcommand\cE{\mathcal{E}}
\newcommand\cF{\mathcal{F}}

\newcommand\cH{\mathcal{H}}
\newcommand\cI{\mathcal{I}}

\newcommand\cK{\mathcal{K}}
\newcommand\cL{\mathcal{L}}
\newcommand\cM{\mathcal{M}}
\newcommand\cN{\mathcal{N}}

\newcommand\C{\mathbb{C}}

\newcommand\fm{\mathfrak{m}}

\definecolor{cardinal}{rgb}{0.6,0,0}
\definecolor{darkgreen}{rgb}{0,0.5,0}
\definecolor{golden}{rgb}{0.92, 0.7, 0}
\definecolor{midnight}{rgb}{0, 0, 0.5}
\definecolor{darkblue}{rgb}{0.2, 0, 0.8}

\allowdisplaybreaks[2] 

\renewcommand{\date}{\dedicated}

\title{The Twisted Index and Topological Saddles}
\author[a]{Mathew Bullimore,}
\author[a,b]{Andrea E. V. Ferrari,}
\author[b]{Heeyeon Kim,}
\author[a]{Guangyu Xu}
\affiliation[a]{Department of Mathematical Sciences, Durham University, \\ Stockton Road, Durham DH1 3LE, UK}
\affiliation[b]{Mathematical Institute, University of Oxford, \\
Woodstock Road, Oxford, OX2 6GG, UK}
\emailAdd{mathew.r.bullimore@durham.ac.uk}
\emailAdd{andrea.e.v.ferrari@gmail.com}
\emailAdd{kimh@maths.ox.ac.uk}
\emailAdd{guangyu.xu@durham.ac.uk}
\date{\today}

\abstract{
The twisted index of 3d $\cN=2$ gauge theories on $S^1 \times \Sigma$ has an algebro-geometric interpretation as the Witten index of an effective supersymmetric quantum mechanics. In this paper, we consider the contributions to the supersymmetric quantum mechanics from topological saddle points in supersymmetric localisation of abelian gauge theories. Topological saddles are configurations where the matter fields vanish and the gauge symmetry is unbroken, which exist for non-vanishing effective Chern-Simons levels. We compute the contributions to the twisted index from both topological and vortex-like saddles points and show that their combination recovers the Jeffrey-Kirwan residue prescription for the twisted index and its wall-crossing. 
}

\begin{document}
\maketitle
\flushbottom
\section{Introduction}

\subsection{A Quick Summary}

A powerful feature of supersymmetric localisation is the ability to introduce exact deformations and scaling limits that lead to different mathematical representations of the same partition function. 
In the case of 3d $\cN=2$ supersymmetric gauge theories, the localisation schemes used in the literature fall into two broad classes: 
\begin{itemize}
\item In Coulomb branch localisation, the path integral localises onto configurations where the vectormultiplet scalar is non-zero and the gauge group is broken to a maximal torus. This expresses the partition function as a contour integral of special functions, for example, as in the first supersymmetric localisation computations on  $S^3$~\cite{Kapustin:2009kz,Hama:2011ea,Hama:2010av}. 
\item In Higgs branch localisation, the path integral localises instead onto configurations solving vortex-like equations. This expresses the path integral in terms of integrals of characteristic classes over moduli spaces of vortices, which may often be reduced to isolated fixed points by turning on mass parameters for flavour symmetries~\cite{Fujitsuka:2013fga,Benini:2013yva}. 
\end{itemize}
In many cases, the individual residues of the contour integral in Coulomb branch localisation reproduce the isolated fixed point contributions in Higgs branch localisation. Further background on  localisation in three dimensions can be found in the reviews~\cite{Willett:2016adv,Closset:2019hyt}. 

In this paper, we focus on the twisted index of 3d $\cN=2$ supersymmetric gauge theories on $S^1 \times \Sigma$, where $\Sigma$ is a closed Riemann surface of genus $g$~\cite{Benini:2015noa,Benini:2016hjo,Closset:2016arn}. This is a rich observable amenable to supersymmetric localisation and has important applications to exact microstate counting for supersymmetric black holes in AdS$_4$~\cite{Benini:2015eyy,Hosseini:2016tor,Benini:2016rke,Zaffaroni:2019dhb} and the Bethe/gauge correspondence \cite{Nekrasov:2014xaa,Chung:2016lrm,Kanno:2018qbn}.
The aim is to expand upon and generalise the geometric interpretation of the twisted index as the Witten index of an effective supersymmetric quantum mechanics~\cite{Bullimore:2018yyb,Bullimore:2018jlp,Bullimore2019}. 

The route to such a geometric interpretation of the twisted index is through a Higgs branch localisation scheme and for large classes of theories the twisted index is a generating function of enumerative invariants associated to moduli spaces of vortices on the Riemann surface $\Sigma$. However, Higgs branch localisation of the twisted index of a generic 3d $\cN=2$ gauge theory leads to additional saddle points beyond vortex configurations. For a $U(1)$ gauge theory, these additional saddle points are characterised by
\begin{enumerate}
\item All chiral multiplets vanish $\Phi = 0$;
\item The vectormultiplet scalar is fixed $\sigma = \sigma_0$;
\item The gauge symmetry is unbroken.
\end{enumerate}
Such ``topological'' saddle points exist whenever the effective $U(1)$ Chern-Simons level in asymptotic regions of the Coulomb branch is non-zero. We compute the contributions of topological saddle points to the twisted index and show that they reproduce the contributions from residues at infinity in Coulomb branch localisation. These contributions are crucial for the consistency of the geometric interpretation of the twisted index and wall-crossing phenomena studied in~\cite{Bullimore2019} in more general classes of theories.

\subsection{Some More Details}

Let us first summarise the Coulomb branch localisation scheme for the twisted index of 3d $\cN=2$ gauge theories on $S^1\times \Sigma$, which was introduced for $g = 0$ in~\cite{Benini:2015noa} and extended to $g >0$ in~\cite{Benini:2016hjo,Closset:2016arn}. For a $ U(1)$ gauge theory this leads to a representation of the twisted index as a contour integral
\be
\cI = \sum_{\fm \in \mathbb{Z}} q^\mathfrak{m} \int_{\Gamma} \frac{dx}{2\pi i x} \, \cI(x,\fm) \, ,
\label{eq:coulomb-intro}
\ee
where each summand is the contribution from configurations with $\fm$ units of flux on $\Sigma$~\footnote{We assume there are no monopole operators in the superpotential so there is a $U(1)$ topological symmetry}. The integrand receives contributions from a 1-loop determinant and an integral over gaugino zero modes. The contour is a Jeffrey-Kirwan residue prescription, building on computations of the elliptic genus of 2d $\cN=(2,2)$ gauge theories~\cite{Benini:2013nda,Benini:2013xpa}. However, a novel feature in three dimensions is the existence of poles at $x^{\pm1} \to 0$. The contour prescription for these poles is determined by the effective Chern-Simons levels $k^\pm_{\text{eff}}$ in asymptotic regions of the Coulomb branch.

Here we introduce an exact deformation of the lagrangian depending on a real parameter $\tau$. This can be understood as a 1d Fayet-Iliopoulos parameter from the perspective of supersymmetric quantum mechanics on $S^1$. In Coulomb branch localisation, it modifies the contour prescription $\Gamma$ for the poles at $x^{\pm1}\to0$ in~\eqref{eq:coulomb-intro} when the corresponding effective Chern-Simons level vanishes, $k_\text{eff}^\pm = 0$. This has two important consequences:
\begin{itemize}
\item Away from walls in the parameter space of $\tau$, the contour $\Gamma$ is always well-defined for each individual flux $\fm$, before summing over $\fm \in \mathbb{Z}$. This feature is necessary for a hamiltonian interpretation of the twisted index as counting supersymmetric ground states and in particular for compatibility with any Higgs branch localisation scheme. 
\item It leads to interesting wall-crossing phenomenon in $\tau$~\cite{Bullimore2019}, via the same mechanism as supersymmetric quantum mechanics~\cite{Hori:2014tda}. 
\end{itemize}

Indeed, in the presence of the 1d Fayet-Iliopoulos parameter $\tau$, it is possible to consider an alternative scaling limit in the path integral that leads to a Higgs branch localisation scheme. This provides a representation of the twisted index in the form
\be
\cI = \sum_{\mathfrak{m} \in \mathbb{Z}} q^{\mathfrak{m}} \int \hat{\text{A}} (\cM_\mathfrak{m}) \, \text{ch}(\mathcal{F}) \, ,
\label{eq:geometric-intro}
\ee
where $\cM_\fm$ denotes the moduli space of supersymmetric saddle points with flux $\fm$ and $\text{ch}(\cF)$ is the contribution of massive fluctuations and Chern-Simons terms. Each summand in~\eqref{eq:geometric-intro} is the contribution from an effective supersymmetric quantum mechanics in the topologically distinct sector labelled by the flux $\fm$. The moduli space $\cM_\fm$ in general has contributions from two types of saddle points:
\begin{enumerate}
\item {\bf Vortex Saddles} \\ 
Vortex saddles are solutions of abelian vortex equations on $\Sigma$ depending on $\tau$. In the presence of generic mass parameters for flavour symmetries, their contribution to the moduli space $\cM_\fm$ has a concrete description as a disjoint union of symmetric products of $\Sigma$. We show that their contribution to~\eqref{eq:geometric-intro} reproduces residues of the contour integral~\eqref{eq:coulomb-intro} at poles at finite values of $x$.
\item {\bf Topological Saddles} \\
Topological saddles are configurations where all chiral multiplets vanish, $\sigma$ has a fixed expectation value, and the $U(1)$ gauge symmetry is unbroken. Topological saddles only exist if an effective Chern-Simons level is non-vanishing, $k_{\text{eff}}^\pm \neq 0$. Their contribution to the moduli space $\cM_\fm$ is roughly the Picard variety parametrising holomorphic line bundles on $\Sigma$ of degree $\fm$. We show that their contribution to~\eqref{eq:geometric-intro}  reproduces residues of~\eqref{eq:coulomb-intro} at poles at $x^{\pm1} \to 0$.
\end{enumerate}
The existence of both vortex and topological saddle points depends on $\tau$ and the moduli space $\cM_\fm$ may jump as this parameter crosses walls proportional to the flux $\fm$. Taking this into account, we show that equation~\eqref{eq:geometric-intro} exactly reproduces the Coulomb branch residue prescription in the presence of $\tau$ for a broad class of $U(1)$ gauge theories.

Finally, the twisted index has a hamiltonian interpretation as a Witten index
\be
\cI = \sum_{\fm \in \mathbb{Z}} q^\fm \mathrm{Tr}_{\cH_\fm}(-1)^F\, ,
\ee
where $\cH_\fm$ is the space of supersymmetric ground states with flux $\fm$ on $\Sigma$. It is therefore natural to identify the supersymmetric ground states with some form of cohomology of the moduli spaces of vortex and topological saddle points contributing to $\cM_\fm$~\cite{Bullimore:2018yyb}. However, the twisted index exhibits cancelations between contributions from residues at finite $x$ and $x^\pm \to \infty$. This indicates the presence of instanton corrections between perturbative ground states associated to vortex and topological vacua. We hope to return to this phenomenon in the future.

\subsection{Outline}

The outline of the paper is as follows. In section~\ref{sec:background} we review the Coulomb branch localisation of the twisted index of $U(1)$ gauge theories. In section~\ref{sec:geometric} we introduce an alternative Higgs branch localisation scheme and discuss general features of vortex and topological saddles. In sections~\ref{sec:vortex} and ~\ref{sec:topological} we evaluate the contribution of vortex and topological saddles respectively to the twisted index. In section~\ref{sec:examples} we evaluate these contributions explicitly in examples with a single chiral multiplet. Finally, in section~\ref{sec:su2} we perform a preliminary investigation of an $SU(2)$ gauge theory deformed by a 1d Fayet-Iliopoulos parameter for the Cartan subalgebra.

\section{Background}
\label{sec:background}

\subsection{Abelian Theories}
\label{sec:abelian-theories}

We consider a 3d $\mathcal{N}=2$ gauge theory with $G=U(1)$ and $N$ chiral multiplets $\Phi_j$ of charge $Q_j$ and R-charge $r_j \in \mathbb{Z}$. We will restrict attention to the cases where $|Q_j| =1$. We introduce a supersymmetric Chern-Simons term at level $k$ for the gauge group and a mixed gauge-R symmetry Chern-Simons term $k_{\text{R}}$. The quantisation condition requires
\bea
& k + \frac{1}{2} \sum_{j = 1}^N  Q_j^2 \, \in \mathbb{Z} \,, \label{eq:k-constraint}\\
& k_{\text{R}} + \frac{1}{2} \sum_{j=1}^N Q_j(r_j-1) \, \in \mathbb{Z} \,.
\eea
The latter condition implies that the R-symmetry line bundle that we use for the topological twisting is well-defined. 
In this paper, we set the superpotential to vanish.

There is a global topological symmetry $U(1)_\text{t}$ with associated real Fayet-Iliopoulos parameter $\zeta \in \mathbb{R}$. In some cases, this is enhanced to a non-abelian symmetry in the infrared. In addition, there is a global flavour symmetry $G_\text{f}$ with maximal torus
\begin{equation}
T_{\text{f}} \cong \Big[ \prod_{j=1}^N U(1)_j  \Big] / U(1) \,,
\end{equation}
where $U(1)_j$ rotates $\Phi_j$ with charge $+1$ and the quotient is by the gauge group. Correspondingly, we introduce real mass parameters $m_j$ such that the total mass of $\Phi_j$ is $m_j +Q_j \sigma$ where $\sigma$ is the real scalar in the vectormultiplet. The real masses are defined up to a constant shift $m_j \to m_j + Q_j c$, which  can be absorbed by $\sigma \to \sigma - c$. 

Integrating out chiral multiplets in the presence of generic real masses generates effective supersymmetric gauge and mixed gauge-R symmetry Chern-Simons levels
\bea
k_\text{eff}(\sigma) & = k + \frac{1}{2} \sum_{j=1}^N Q_j^2 \, \sign(m_j + Q_j \sigma) \, , \\
k_{R,\text{eff}}(\sigma) & = k_R + \frac{1}{2} \sum_{j=1}^N Q_j(r_j-1) \, \sign(m_j + Q_j \sigma) 
\label{eq:CS-eff}\, ,
\eea
which are piece-wise constant in $\sigma$. The quantisation condition~\eqref{eq:k-constraint} ensures that the constant values are integers. It is also useful to define 
\bea
k^\pm_{\text{eff}} := \lim_{\sigma \to \pm \infty} k_{\text{eff}} = k \pm \frac{1}{2} \sum_{j=1}^N Q_j |Q_j| \, ,
\eea
which controls the gauge charges of monopole operators with $U(1)_\text{t}$ topological charge $\pm1$. Similarly, the asymptotic values of the effective mixed Chern-Simons level controls of the R-charge of the same monopole operators.

\subsection{The Twisted Index}
\label{sec:twisted-index}

Following~\cite{Benini:2016hjo,Benini:2015noa,Closset:2016arn}, we consider the twisted index on $S^1 \times \Sigma$ with a closed orientable Riemann surface $\Sigma$ of genus $g$. The twist is performed using the unbroken R-symmetry, which preserves an $\mathcal{N}=(0,2)$ quantum mechanics on $S^1$ with a pair of supercharges $Q, \bar{Q}$. 

The twisted index has a hamiltonian interpretation as counting supersymmetric ground states on $S^1 \times \Sigma$ annihilated by $Q$, $\bar Q$. The space of supersymmetric ground states $\cH$ transforms as a representation space of the global symmetry $U(1)_\text{t} \times G_\text{f}$ and the index is
\be
\cI = \text{Tr}_{\cH}(-1)^F q^{J_\text{t}} \prod_{j=1}^N y_j^{J_j} \, ,
\label{eq:index-def}
\ee
where $J_\text{t}$ and $J_j$ denote the Cartan generators of $U(1)_\text{t}$ and $U(1)_j$ respectively, and $q$ and $y_j$ are the fugacities. A basic assumption is that $\cH$ is locally finitely graded, meaning that the coefficient of a given monomial in $q$ and $y_j$ is a finite integer. 

\subsection{Supersymmetric Lagrangians}
\label{sec:lagrangians}

The twisted index can also be computed using supersymmetric localisation applied to the path integral construction~\cite{Benini:2016hjo,Benini:2015noa,Closset:2016arn}. We now briefly summarise the various lagrangians used in supersymmetric localisation.
 
First, the vectormultiplet Yang-Mills lagrangian $L_\text{YM}$ and the chiral multiplet lagrangian $L_{\Phi}$ are exact with respect to both supercharges $Q, \bar{Q}$. In supersymmetric localisation, their coefficients are typically sent to infinity so that the saddle point approximation is exact. 

The Fayet-Iliopoulos and mass parameter lagrangians $L_\text{FI}$ and $L_\text{m}$ are not exact and the twisted index depends on these parameters. These parameters are naturally complexified by Wilson lines around $S^1$ for the associated global symmetries. In particular, we can identify the fugacities in the hamiltonian definition~\eqref{eq:index-def} of the twisted index as
\be
q=e^{-2\pi \beta ( \zeta+ i A_\text{t}) }\, , \quad y = e^{-2\pi \beta (m+iA_\text{f})} \, ,
\ee
where $A_{\text{t}}$ and $A_{\text{f}}$ denote the constant background connections for $U(1)_{\text{t}}$ and $T_{\text{f}}$ respectively. The twisted index is a meromorphic function of $q$ and $y$ with poles at loci where non-compact massless degrees of freedom appear and the spectrum is no longer gapped. 

The supersymmetric Chern-Simons lagrangian $L_\text{CS}$ is not exact and the index depends on the level $k$.

Following~\cite{Bullimore2019}, we introduce an additional exact term
\begin{equation}
	L_{\tau} = \frac{i \tau}{2} (Q+\bar{Q})(\lambda+\bar{\lambda}) = -i \tau D_{\text{1d}}
\label{eq:1d-FI}\, ,
\end{equation}
where
\be
D_\text{1d} := D - 2 F_{1\bar 1}
\label{eq:1d-aux}
\ee
and $\tau$ is a real parameter valued in the Lie algebra of the topological symmetry $U(1)_{\text{t}}$. This is interpreted as a 1d Fayet-Iliopoulos parameter since the combination $D_{\text{1d}}$ is the auxiliary field in the $\cN=(0,2)$ quantum mechanics vectormultiplet. Unlike the 3d Fayet-Iliopoulos parameter $\zeta$, it cannot be complexified. As this parameter is real and exact, the twisted index depends in a piecewise constant manner on $\tau$ but may jump accross walls.

\subsection{Contour Integral Formula}
\label{sec:contour}

We now briefly review the derivation of the contour integral formula for the twisted index using supersymmetric localisation. 

First, the localisation scheme used in~\cite{Benini:2015noa,Benini:2016hjo,Closset:2016arn} starts from the lagragian
\begin{equation}
	L = \frac{1}{e^2} L_{\text{YM}} + \frac{1}{g^2} L_{\Phi} + L_{\text{CS}} + L_{\text{FI}} \,.
\label{eq:coulomb-lagrangian}
\end{equation}
Schematically, localisation is done by sending $e^2 \to 0$ and $g^2\rightarrow 0$ in a careful way as described in \cite{Benini:2015noa,Benini:2016hjo,Closset:2016arn}. Following computations of the elliptic genus of supersymmetric gauge theories in two dimensions~\cite{Benini:2013nda,Benini:2013xpa}, a more precise analysis leads to a Jeffrey-Kirwan residue integral.

The contour integral takes the form

\begin{equation}
	\mathcal{I} = \sum_{\mathfrak{m}\in \mathbb{Z}} q^\mathfrak{m}\frac{1}{2 \pi i} \oint_{\mathcal{C}} \frac{\dd x}{x} \, H(x)^g Z(x,\mathfrak{m}) \,,
\label{eq:partition-function-contour-integral}
\end{equation}
where the contour $\cC$ is in the complexified maximal torus of $G=U(1)$ parameterised by $x$ and the summation is over the magnetic flux $\mathfrak{m}\in \mathbb{Z}$. The integrand is constructed from the supersymmetric Chern-Simons and 1-loop contributions
\begin{equation}
	Z(x,\mathfrak{m}) =  x^{k \mathfrak{m}}  x^{(g-1)k_R}  \prod_{i=1}^N \left[ \frac{(x^{Q_i} y_i)^{\frac{1}{2}}}{1-x^{Q_i} y_i} \right]^{Q_i \mathfrak{m}+ (g-1)(r_i-1)} 
\label{eq:1l}
\end{equation}
and the hessian factor
\bea
H(x) 
 = k + \sum_{j=1}^N  Q_j^2 \left(\frac{1}{2} + \frac{x^{Q_j} y_j}{1 - x^{Q_j} y_j}\right)\, ,
\label{eq:hessian-factor}
\eea
which arises from integration over gaugino zero modes. 

The contour is given explicitly by
\be
\frac{1}{2\pi i} \oint_{\mathcal{C}} \frac{\dd x}{x} = \sum_{x_*} \JK{x_*} \left(Q_{*},\eta\right)  \frac{\dd x}{x} \, ,
\ee
where 
\begin{equation}
	\JK{0} (Q,\eta) \frac{\dd x}{x} := \Theta (Q \eta ) \, \sign(Q) 
\label{eq:JK}
\end{equation}
and $\eta \neq 0$ is an auxiliary real parameter. The sum is over poles of the integrand and $Q_*$ denotes the Jeffrey-Kirwan charge associated to a pole at $x = x_*$. The poles at solutions of $x^{Q_i} y_i  = 1$ arise from the elementary chiral multiplet $\Phi_i$ and their Jeffrey-Kirwan charge is simply the gauge charge $Q_i$. The Jeffrey-Kirwan charges assigned to the poles at $x = 0$ and $x = \infty$ are 
\bea
x = 0  : \quad Q_+  & = -k_{\text{eff}}^+\, \, ,\\
x = \infty : \quad Q_- & = +k_{\text{eff}}^-\, ,
\label{eq:mono-charge-1}
\eea
which are the gauge charges of monopole operators of $U(1)_t$ charges $\pm1$.

This result is manifestly independent of the auxiliary parameter $\eta$ if all Jeffrey-Kirwan charges are non-vanishing. However, if a monopole operator is gauge neutral, when $Q_0 = 0$ or $Q_\infty = 0$, then the Jeffrey-Kirwan residue operation requires further specification. In such cases, \cite{Benini:2015noa,Benini:2016hjo,Closset:2016arn} introduce an additional term in the lagrangian as a regulator with the result that the residue at $x = 0$ or $x=\infty$ should not be taken in such cases. This leads to a meaningful result that is independent of $\eta$ after summing over magnetic flux $\mathfrak{m} \in \mathbb{Z}$.

In reference~\cite{Bullimore2019}, the 1d Fayet-Iliopoulos parameter $\tau$ was introduced to ensure a meaningful result for each individual flux $\mathfrak{m}\in \mathbb{Z}$. This feature is necessary if we want to unambiguously interpret the coefficient of $q^\mathfrak{m}$ as counting the supersymmetric ground states with $U(1)_t$ charge $\mathfrak{m}$, as in the hamiltonian definition~\eqref{eq:index-def}. The starting point is now the lagrangian
\begin{equation}
	L = \frac{1}{t^2}\left(\frac{1}{e^2} L_{\text{YM}} + L_{\tau}\right) + \frac{1}{g^2} L_{\Phi} + L_{\text{CS}} + L_{\text{FI}} \,.
\label{eq:modified-coulomb-lagrangian}
\end{equation}
and the localisation proceeds as in supersymmetric quantum mechanics by taking the limit $t^2 \rightarrow 0$ with $e^2$ finite~\cite{Hori:2014tda}. This leads to an identical contour integral formula but with a different assignment of Jeffrey-Kirwan charges to the poles.

For the poles associated to chiral multiplets $\Phi_i$ the Jeffrey-Kirwan charge is again $Q_i$. For the poles associated to monopole operators, the Jeffrey-Kirwan charges are now assigned according to
\begin{subequations}
\label{eq:JK-charges-boundary}
\begin{align}
\label{eq:JK-charges-0}
	Q_{+} &= 
	\begin{dcases}
    	- k_\eff^+ \quad &\text{if} \quad k_\eff^+  \neq 0 \\
    	\mathfrak{m}- \tau' \quad &\text{otherwise} 
  	\end{dcases}\,, \\
 \label{eq:JK-charges-inf}
	Q_{-} &= 
	\begin{dcases}
    	+ k_\eff^-\quad &\text{if} \quad k_\eff^- \neq 0 \\
    	\mathfrak{m}- \tau' \quad &\text{otherwise} 
  	\end{dcases}\,,
\end{align}
\end{subequations}
where $\tau$ is re-scaled as
\begin{equation}
	\tau':=\frac{e^2 \mathrm{vol}(\Sigma)}{2 \pi} \tau\,.
\label{eq:normalised-1d-FI}
\end{equation}
The contribution from each magnetic flux $\mathfrak{m}\in \mathbb{Z}$ is now separately independent of the auxiliary parameter $\eta$ provided $\tau' \neq \mathfrak{m}$. However, the twisted index may now jump accross the wall $\tau ' = \mathfrak{m}$ according to
\be
\label{eq:wc}
\cI(\tau' = \mathfrak{m}+\ep) - \cI(\tau' = \mathfrak{m} - \ep) = q^\mathfrak{m}\left[ \, \delta_{k_{\text{eff}}^+,0} \; \underset{x=0}{\mathrm{Res}}+ \delta_{k_{\text{eff}}^-,0} \; \underset{x=\infty}{\mathrm{Res}} \,\right]\frac{\dd x}{x} \, H(x)^g Z(x,\mathfrak{m}) \, .
\ee
where $\ep  \to 0^+$. 

In what follows, we therefore require $\tau' \notin \mathbb{Z}$. 
This ensures the Jeffrey-Kirwan charges are always non-vanishing and the contribution to the twisted index from each flux $\fm \in \mathbb{Z}$ is meaningful.

Finally, the contour prescription used in~\cite{Benini:2015noa,Benini:2016hjo,Closset:2016arn} is recovered by sending $\tau' \to +\infty$ with $\eta >0$ or $\tau' \to - \infty$ with $\eta <0$. That this is independent of the auxiliary parameter $\eta$ is equivalent to the statement that sum of~\eqref{eq:wc} over $\fm \in \mathbb{Z}$ is proportional to a formal delta function at $q = 1$.

\section{Alternative Localisation Scheme}
\label{sec:geometric}

\subsection{Motivation}

The hamiltonian definition of the twisted index~\eqref{eq:index-def} can be viewed as the Witten index of an effective $\cN=(0,2)$ supersymmetric quantum mechanics obtained by twisting on $S^1 \times \Sigma$. Based on the general structure of supersymmetric quantum mechanics of this type we can expect a geometric construction of the twisted index in the form
\be
\cI = \sum_{\mathfrak{m} \in \mathbb{Z}} q^{\mathfrak{m}} \int \hat{\text{A}} (\mathfrak{M}_\mathfrak{m}) \, \text{ch}(\mathcal{\mathcal{E}_\mathfrak{m}}) \, .
\label{eq:geometric-contruction}
\ee
In such an expression, $\mathfrak{M}_\mathfrak{m}$ denotes a moduli space parametrising saddle points of the localised path integral with magnetic flux $\mathfrak{m} \in \mathbb{Z}$, while $\mathcal{E}_\mathfrak{m}$ is schematically a complex of vector bundles arising from the massive fluctuations of chiral multiplets and supersymmetric Chern-Simons terms. The integral should be understood equivariantly with respect to the flavour symmetry $T_{\text{f}}$, leading to the dependence on the parameters $y_i$.

For such an interpretation to be meaningful, it is necessary for the contribution to the twisted index from each individual flux $\mathfrak{m} \in \mathbb{Z}$ to be unambiguous. This necessitates the introduction of the 1d Fayet-Iliopoulos parameter $\tau$. The  wall-crossing phenomena in $\tau$ are then reflected in jumps in the  structure of the moduli spaces $\mathfrak{M}_{\mathfrak{m}}$ and complexes $\mathcal{E}_\mathfrak{m}$.

This general expectation was verified in previous work~\cite{Bullimore:2018jlp,Bullimore2019} for a special class of theories (for example those with $\cN=4$ supersymmetry) where for generic $\tau' \neq \mathfrak{m}$ the moduli spaces $\mathfrak{M}_\mathfrak{m}$ exclusively parametrise vortex-like configurations on $\Sigma$ where the gauge group is completely broken. The purpose of this paper is to extend the geometric interpretation to theories with ``topological" saddle points, where there is an unbroken gauge symmetry and the moduli spaces $\mathfrak{M}_{\fm}$ must be described as quotient stacks.

There is an important distinction between saddle points where the unbroken gauge symmetry is the whole $G = U(1)$ or a discrete subgroup. The latter involves a relatively mild extension of~\cite{Bullimore2019} to deal with moduli spaces with orbifold singularities and our constraint $|Q_i| = 1$ is designed to avoid such cases. We therefore consider theories with topological saddle points where $G = U(1)$ is fully unbroken and the moduli space $\mathfrak{M}_{\mathfrak{m}}$ has a component that is the Picard stack parametrising degree $\mathfrak{m}$ holomorphic line bundles on $\Sigma$.

\subsection{Localising Action}
\label{subsec:localising-action-alternaative}

To arrive at such a geometric interpretation we introduce an alternative supersymmetric localisation for the twisted index~\cite{Bullimore2019}, which is similar to the Higgs branch localisation schemes for 2d $\cN=(2,2)$ theories~\cite{Benini:2012ui,Doroud:2012xw,Closset:2015rna} and for 3d $\cN=2$ theories~\cite{Fujitsuka:2013fga,Benini:2013yva}. However, in addition to the usual vortex-like saddle points, here there will be additional topological saddle points where the matter fields vanish and the gauge group is unbroken.

The starting point is to consider the same lagrangian from equation~\eqref{eq:modified-coulomb-lagrangian} including the 1d Fayet-Iliopoulos parameter $\tau$ but first set $g=t$ to obtain 
\begin{equation}
	L = \frac{1}{t^2}\left(\frac{1}{e^2} L_{\text{YM}} + L_{\tau} + L_{\Phi}\right) + L_{\text{CS}} + L_{\text{FI}} \,.
\label{eq:higgs-lagrangian}
\end{equation}
The second step is to consider the limit $t^2 \rightarrow 0$ while keeping $e^2$ finite. The supersymmetric saddle points are then solutions to the following set of equations,
\begin{subequations}
\label{eq:higgs-localisation-vacuum}
\begin{equation}
	\frac{1}{e^2} * F_A + \sum_{j =1}^N  Q_j |\phi_j|^2 - t^2\sigma \frac{k_\eff(\sigma) }{2\pi} - \tau = 0 \,,
\label{eq:higgs-vortex-eqaution}
\end{equation}
\begin{equation}
	\dd_A \sigma =0\,, \quad \bar{\partial}_A \phi_i =0\,,
\label{eq:higgs-sigma-phi}
\end{equation}
\begin{equation}
	(m_i + Q_i \sigma) \phi_i = 0 \quad \forall \, i  = 1,\ldots, N \,,
\label{eq:higgs-m-phi}
\end{equation}
\end{subequations}
where $F_A$ is the curvature of the gauge connection $A$, and $*$ is the Hodge star operator on $\Sigma$. In writing these equations, $\phi_j$ should be understood as a section of $K_\Sigma^{r_j/2} \otimes L^{Q_j}$ where $K_\Sigma$ is the canonical bundle on $\Sigma$ and $L$ is the holomorphic gauge bundle on $\Sigma$.

Note that the dependence on the 3d Fayet-Iliopoulos parameter $\xi$ has dropped out but the equations depend critically on the 1d Fayet-Iliopoulos parameter $\tau$. We keep the contribution proportional to the effective Chern-Simons term in the limit $t^2 \to 0$ to capture potential saddle points where $|\sigma| \to \infty$ with $\sigma_0 : = t^2\sigma$ finite.

\subsection{Saddle Points}

The solutions to equations \eqref{eq:higgs-localisation-vacuum} fall into topologically distinct sectors labelled by the flux
\be
\mathfrak{m} := \frac{1}{2\pi}\int_\Sigma F_A \in \mathbb{Z} \, .
\ee
A constraint on the existence of saddle points with a given flux $\mathfrak{m}$ is found by integrating equation~\eqref{eq:higgs-vortex-eqaution} over the Riemann surface $\Sigma$ to give
\begin{equation}
\left(\tau' - \mathfrak{m}\right) +\frac{e^2\vol(\Sigma)}{4\pi^2}  t^2 \sigma k_{\eff}(\sigma) = \sum_{j=1}^N  Q_j \, \norm{\phi_j}^2 \,,
\label{eq:integrated-vortex-equation}
\end{equation}
where 
\be
\norm{\phi_j}^2 := \frac{e^2}{2\pi}\int_\Sigma \bar\phi_j \wedge * \, \phi_j
\ee
is a positive definite inner product on sections of $K^{r_j}_\Sigma \otimes L^{Q_j}$ and $\tau'=\frac{e^2 \mathrm{vol}(\Sigma)}{2 \pi} \tau$ is the normalised 1d Fayet-Iliopoulos parameter defined in \eqref{eq:normalised-1d-FI}.

Assuming the 1d Fayet-Iliopoulos parameter is generic, meaning $\tau' \neq \mathfrak{m}$, there are two classes of solutions with a given magnetic flux $\mathfrak{m} \in \mathbb{Z}$. They can be described as follows:
\begin{enumerate}
	\item {\bf Vortex Saddles} 
	
	Vortex saddle points are solutions where $\sigma$ remains finite in the limit $t^2 \to 0$ and the term proportional the effective Chern-Simon level $k_{\text{eff}}$ in equation \eqref{eq:higgs-vortex-eqaution} can be ignored. The remaining equations are the vortex equations
\be
\frac{1}{e^2} * F_A + \sum_{j =1}^N  Q_j |\phi_j|^2  = \tau \, , \quad \bar\partial_A \phi_i = 0 \, , \quad (m_i + Q_i \sigma) \phi_i = 0 \, .
\label{eq:vortex-eq-1}
\ee
For generic mass parameters $m_1,\ldots,m_N$, the space of solutions decomposes as a disjoint union of components where a single $\phi_i$ is non-vanishing and $\sigma = -m_i /Q_i$. From the constraint \eqref{eq:integrated-vortex-equation}, a component of the moduli space where $\phi_i$ is non-vanishing exists if
\be
\sign(\tau'-\fm) = \sign(Q_i) \, .
\ee
	\item {\bf Topological Saddles}
	
	Topological saddle points are solutions where $|\sigma| \to \infty$ in the limit $t^2 \to 0$, such that the combination $\sigma_0:=t^2 \sigma$ remains finite and has a unique non-vanishing solution. This requires $\phi_j = 0$ for all $j=1,\ldots,N$ and therefore the constraint \eqref{eq:integrated-vortex-equation} becomes 
	\begin{equation}
		\tau' - \mathfrak{m} = - \frac{e^2\vol(\Sigma)}{4\pi^2} \sigma_0 k_{\eff}^\pm  \,.
	\label{eq:vortex-topological}
	\end{equation}		
A unique solution with $\pm\sigma_0 >0$ exists if $k_{\eff}^\pm \neq 0$ and 
\be
\sign(\tau'-\mathfrak{m}) = \sign(Q_\pm) \, .
\ee

\end{enumerate}
In addition, if $k_{\eff}^\pm = 0$ then a non-compact Coulomb branch parametrised by $\pm\sigma_0 >0$ appears at $\tau'=\mathfrak{m}$, which is responsible for the wall-crossing phenomena in equation~\eqref{eq:wc}. These three classes are analogous to the trichotomy of flat space supersymmetric vacua in~\cite{Intriligator:2013lca}.

If we align the auxiliary parameter
\be
\sign(\tau'-\fm) = \sign(\eta) \, ,
\ee
components of the moduli space of saddles with flux $\fm$ are in one-to-one correspondence with the poles selected by the contour prescription in section~\ref{sec:background}. There is a component of the vortex moduli space with $\phi_i \neq 0$ when the pole at $x^{Q_i}y_i = 1$ is selected. Similarly, there is a topological saddle point with $\pm\sigma_0 >0$ whenever the residue at $x^{\pm1} \to 0$ is selected. The purpose of sections~\ref{sec:vortex} and~\ref{sec:topological} is to reproduce the residues at these poles.

\section{Vortex Saddles}
\label{sec:vortex}

\subsection{Moduli Space}
\label{sec:vortex-moduli}

The moduli space of vortex saddle points consists of solutions to 
\be
\begin{gathered}
\frac{1}{e^2} * F_A + \sum_{j =1}^N  Q_j |\phi_j|^2  = \tau \, , \\ 
\bar\partial_A \phi_j = 0 \, ,\quad (m_j + Q_j \sigma) \phi_j = 0 \, ,
\end{gathered}
\label{eq:vortex-eq-2}
\ee
for all $j = 1,\ldots,N$, modulo gauge transformations. The moduli space is a disjoint union of topologically distinct components $\cM_\mathfrak{m}$ labelled by the magnetic flux $\mathfrak{m}\in \mathbb{Z}$. The entire moduli space is realised as an infinite-dimensional K\"ahler quotient and under our assumptions each component $\cM_\mathfrak{m}$ is a finite-dimensional smooth K\"ahler manifold.
 
For generic mass parameters $m_i$, the moduli space further decomposes as a disjoint union of components $\cM_{\mathfrak{m},i}$ where a single chiral multiplet $\phi_i$ is non-vanishing and $\sigma = - m_i / Q_i$. Each component parametrises solutions to the abelian vortex equations
\be
\frac{1}{e^2} * F_A + Q_i |\phi_i|^2  = \tau \, ,  \quad \bar\partial_A \phi_i = 0  \, ,
\label{eq:vortex-eq-part}
\ee
where $\phi_i$ transforms as a section of $K_\Sigma^{r_i/2} \otimes L^{Q_i}$. A small modification of the standard analysis applies and each component is either a symmetric product of the curve $\Sigma$ or empty,
\bea
Q_i = +1: \quad \cM_{\mathfrak{m},i} = \begin{cases}
\Sigma_{d_i} & \text{if}  \quad \tau' > \mathfrak{m} \\
\emptyset & \text{if}  \quad \tau' < \mathfrak{m}
\end{cases} \\
Q_i = -1: \quad \cM_{\mathfrak{m},i} = \begin{cases}
\emptyset & \text{if}  \quad \tau' > \mathfrak{m} \\
\Sigma_{d_i} & \text{if}  \quad \tau' < \mathfrak{m}
\end{cases}
\label{eq:reduced-vortex}
\eea
where
\be
d_i : = Q_i \mathfrak{m} + r_i(g-1)
\ee
is the degree of $K_\Sigma^{r_i/2} \otimes L^{Q_i}$ and $\Sigma_d:=\text{Sym}^{d}\Sigma$ with the understanding that this is empty for $d<0$. The symmetric product $\Sigma_{d_i}$ parametrises the positions of the vortices. The assumption $|Q_i| = 1$ is important to get a symmetric product, otherwise the moduli space has orbifold singularities where a discrete gauge subgroup is unbroken.

Note that if the auxiliary parameter $\eta$ is aligned with $\tau' - \mathfrak{m}$, meaning
\be
\text{sign}(\tau' - \mathfrak{m}) = \text{sign}(\eta) \, ,
\ee
then the component $\cM_{\mathfrak{m},i}$ of the moduli space is non-empty whenever the Jeffrey-Kirwan residue prescription selects the pole at $x^{Q_i}y_i = 1$ from the chiral multiplet $\Phi_i$. The task in the remainder of this section is to reproduce the residue at this pole from supersymmetric localisation.

It is useful to use an algebraic description of moduli spaces of abelian vortices in terms of holomorphic pairs. Let us assume $\sign(\tau'-\fm) = \sign(Q_i)$ so that the vortex moduli space $\cM_{\mathfrak{m},i}$ is non-empty. Then the Hitchin-Kobayashi correspondence says that there is an algebraic description parametrising pairs $(L,\phi_i)$ where $L$ is a holomorphic line bundle of degree $\mathfrak{m}$ and $\phi_i$ is a non-zero section of $K_\Sigma^{r_i/2} \otimes L^{Q_i}$. The symmetric product $\Sigma_{d_i}$ in equation~\eqref{eq:reduced-vortex} parametrises the zeros of the section $\phi_i$.

\subsection{Contributions to the Index}
\label{sec:vortex-contributions}

The contribution to the twisted index from a component $\cM_{\fm,i}$ of the vortex moduli space is
\bea
	\cI_{\fm,i} 
	& = \int \hat{A} ( \cM_{\mathfrak{m},i} )  \, \frac{\hat{A}(\mathcal{E}  )}{e( \mathcal{E} )} \text{ch}(\cL^k\otimes \cL^{k_R}_R)) \\
	&  = \int \hat{A} ( \cM_{\mathfrak{m},i} ) \frac{ \text{ch}(\cL^k \otimes \cL^{k_R}_R)}{\text{ch}(\hat \wedge^\bullet \mathcal{E} ) } 
	\label{eq:vortex-general-form} 
\eea
where $\mathcal{E} $ is a perfect complex of sheaves encoding the massive fluctuations of chiral multiplets around vortex configurations, and $\cL$, $\cL_R$ are holomorphic line bundles arising from the gauge and mixed gauge-R symmetry Chern-Simons terms respectively. We omit the labels $\fm,i$ from the bundles for brevity.

This integral should be understood equivariantly with respect to the flavour symmetry $T_{\text{f}}$ with parameters $y_i$. It can be evaluated using intersection theory on symmetric products and converted into a contour integral following~\cite{macdonald1962symmetric,arbarello1985geometry}. This extends a computation performed in~\cite{Bullimore:2018jlp} to a wider class of theories.

For simplicity and to avoid a profusion of signs at intermediate steps in the calculation, we set $Q_i = +1$ with $\tau' > \fm$. A similar computation applies to $Q_i = -1$ and the final result is presented in a uniform way for both cases.

\subsubsection{Tangent Directions}

We first consider the contribution from the tangent directions to $\cM_{\mathfrak{m},i}$, which is the symmetric product $\Sigma_{d_i}$ when $\tau'>\mathfrak{m}$ and otherwise empty.

Let us briefly summarise some notation for the intersection theory on a symmetric product. There are standard generators $\zeta_a$, $\bar\zeta_a \in H^2(\Sigma_d)$ with $a = 1,\ldots g$ and $\eta \in H^2(\Sigma_d)$ arising from cohomology classes on $\Sigma$. We then define $\theta_a := \zeta_a \wedge \bar\zeta_a$ and $\theta := \sum_{a=1}^g \theta_a$. 

From reference~\cite{macdonald1962symmetric}, the Chern character of the tangent bundle is
\bea
\ch( T\Sigma_{d_i}) 
& = (g-1) +((d_i-2g+1)-\theta)e^\eta \\
& = (g-1) + (d_i-2g+1)e^{\eta} + \sum_{a=1}^ge^{\eta-\theta_a} \, .
\label{eq:chern-roots-of-sym}
\eea
From here we can evaluate the $\hat A$-genus as follows
\begin{align}
\hat{A}(\Sigma_{d_i}) 	
	=&  \exp(-\frac{d_i-g+1}{2}\eta + \frac{\theta}{2}) \left(\frac{\eta}{1-e^{-\eta}}\right)^{d_i-2g+1} \prod_{a=1}^g \frac{\eta- \theta_a}{1 - e^{-\eta+ \theta_a}} \nonumber\\
	=&  \exp(-\frac{d_i-g+1}{2}\eta + \frac{\theta}{2}) \left(\frac{\eta}{1-e^{-\eta}}\right)^{d_i-2g+1} \prod_{a=1}^g \frac{\eta- \theta_a}{1 - e^{-\eta}(1+ \theta_a)} \nonumber\\
	=&  \exp(-\frac{d_i-g+1}{2}\eta + \frac{\theta}{2}) \left(\frac{\eta}{1-e^{-\eta}}\right)^{d_i-2g+1} \prod_{a=1}^g \frac{\eta- \theta_a}{(1 - e^{-\eta})-e^{-\eta}  \theta_a} \nonumber\\
	=&  \exp(-\frac{d_i-g+1}{2}\eta + \frac{\theta}{2}) \left(\frac{\eta}{1-e^{-\eta}}\right)^{d_i-2g+1} \prod_{a=1}^g \frac{\eta- \theta_a}{1 - e^{-\eta}}\left(1-\frac{e^{-\eta} }{1 - e^{-\eta}}\theta_a\right)^{-1} \nonumber\\
	=&  \exp(-\frac{d_i-g+1}{2}\eta + \frac{\theta}{2}) \left(\frac{\eta}{1-e^{-\eta}}\right)^{d_i-2g+1} \prod_{a=1}^g \frac{\eta}{1 - e^{-\eta}}\left(1-\frac{ \theta_a}{\eta}+\frac{e^{-\eta} }{1 - e^{-\eta}}\theta_a\right) \nonumber\\
	=&  \exp(-\frac{d_i-g+1}{2}\eta + \frac{\theta}{2}) \left(\frac{\eta}{1-e^{-\eta}}\right)^{d_i-g+1} \prod_{a=1}^g    \exp[\theta_a\left(-\frac{1}{\eta}+\frac{e^{-\eta}}{1 - e^{-\eta}}\right)] \nonumber\\
	=&  \left(\frac{\eta e^{-\eta/2}}{1-e^{-\eta}}\right)^{d_i-g+1}    \exp[\theta\left(\frac{1}{2}-\frac{1}{\eta}+\frac{e^{-\eta}}{1 - e^{-\eta}}\right)] \,,
\label{eq:A-roof-sym}
\end{align}
where we have made repeated use of $\theta_a^2 = 0$ for any $a =1,\ldots,g$.

\subsubsection{Index Bundle}

We now consider the fluctuations of each of the remaining massive chiral multiplets $\Phi_j$ with $j \neq i$ around configurations in $\cM_{\mathfrak{m},i}$. 

At a point $(L,\phi_i)$ on the moduli space $\cM_{\mathfrak{m},i}$, each chiral multiplet $\Phi_j$ with $j \neq i$ generates 1d $\cN=(0,2)$ chiral and Fermi multiplet fluctuations valued in the following vector spaces
\begin{itemize}
\item Chiral multiplets: \quad $E_j^0 := H^0\left(\Sigma,  L^{Q_j} \otimes K_\Sigma^{r_j/2}\right)\,,$
\item Fermi multiplets: \quad $E_j^1 : = H^1\left(\Sigma,L^{Q_j} \otimes K_\Sigma^{r_j/2}\right)\,.$
\end{itemize}
If we move around in the moduli space $\cM_{\mathfrak{m},i}$ the dimensions of these vector spaces may jump. But by the Riemann-Roch theorem the difference of their dimensions is constant,
\bea
\text{dim} \, E_j^0 - \text{dim} \, E_j^1 
& = h^0 (L^{Q_j}\otimes K_{\Sigma}^{r_j/2}) - h^1 (L^{Q_j}\otimes K_{\Sigma}^{r_j/2} )  \\
& = Q_j \mathfrak{m}+ (g-1)(r_j-1) \\
& = d_j -g +1 \, .
\eea
We can therefore formally regard the difference of these vector spaces as the fibre of a holomorphic vector bundle on the moduli space $\cM_{\mathfrak{m},i}$ of rank $d_j-g+1$, or at least this will define a reasonable K-theory class for use in the computation of the twisted index. 

To make this more precise, we recall the construction of the universal bundle on a symmetric product. We consider the pair of projection maps
\begin{equation}
\begin{tikzcd}[column sep=small]
	& {\Sigma_{d_i} \times \Sigma} \arrow[dl,"\pi", swap]\arrow[dr,"p"] & \\ {\Sigma_{d_i}} && {\Sigma}
\end{tikzcd} \,.
\label{eq:sym-projections}
\end{equation}
There is a unique universal line bundle $\cL$ on $\Sigma_{d_i} \times \Sigma$ with the property that its restriction to a point $(L,\phi_i)$ on the symmetric product is the holomorphic line bundle $L$ on $\Sigma$. We also define $\mathcal{K} := p^*K_\Sigma$ to be the pull-back of the canonical bundle on the curve. With this in hand, the fluctuations of $\Phi_j$ transform in a perfect complex of sheaves on $\Sigma_d$ defined by the derived push-forward 
\begin{equation}
	\mathcal{E}^\bullet_j := R^\bullet \pi_*(\mathcal{L}^{Q_j} \otimes \mathcal{K}^{r_j /2}) \, .
\end{equation}
The stalks of $\mathcal{E}^\bullet_j$ over a point $(L,\phi_i)$ on the symmetric product are the vector spaces $E_j^\bullet$.

We can extract the Chern roots of $\mathcal{E}^\bullet_j$ following standard computations~\cite{arbarello1985geometry}. The starting point is the Chern character of the universal bundle
\begin{equation}
	\ch\left(\mathcal{L}^{Q_j}\right) = e^{Q_j \eta} \left(1 + Q_j \mathfrak{m} \,\eta_{\scriptscriptstyle \Sigma} + Q_j \gamma - Q_j^2 \eta_{\scriptscriptstyle \Sigma} \theta \right) 
\end{equation} 
and
\be
\ch(\mathcal{K}^{r_j /2}) =  1+r_j(g-1)\eta_{\scriptscriptstyle \Sigma} \, .
\ee
Here we abuse notation and identify the cohomology classes $\eta$, $\theta$ with their pull-backs by $\pi$. In a similar way, $\eta_\Sigma$ denotes the class of a point on $\Sigma$ and its pullback by $p$. Finally, $\gamma$ is built from the pull-back of $1$-form generators and will not play a role in what follows. 

An application of the Groethendiek-Riemann-Roch theorem to $\pi$ gives
\begin{align}
	\ch(\mathcal{E}^\bullet_j)   &= \pi_*\left\{\ch(\mathcal{L}^{Q_j} \otimes \cK^{r_j/2})  \,  \td( \Sigma_{d_i}) \right\} \nonumber \\
	&= e^{Q_j \eta}\left((d_j -g +1) - Q_j^2 \theta\right) \,.
\end{align}
On vortex saddle points parametrised by the moduli space $\cM_{\mathfrak{m},i}$, the real vectormultiplet scalar is fixed to $\sigma = - m_i$ and the real mass of $\Phi_j$  fluctuations is $m_j - Q_j m_i$. We therefore promote this result to a $T_\text{f}$-equivariant Chern character
\be
\ch(\mathcal{E}^\bullet_j) = z_j e^{Q_j \eta}\left((d_j -g +1) - Q_j^2 \theta\right) \, ,
\ee
where
\be
z_j := y_j / y_i^{Q_j} \, .
\ee
The fluctuations from all the massive chiral multiplets is encoded in
\be
\mathcal{E} = \bigotimes_{j\neq i} \mathcal{E}^\bullet_j \, .
\ee
The equivariant Chern roots have a similar structure to those in \eqref{eq:chern-roots-of-sym} and the contribution of these fluctuations to the twisted index can be evaluated in a similar way,  with the result
\bea
	\frac{\hat{A}(\mathcal{E})}{e(\mathcal{E} )}  
=  \prod_{j\neq i} \, \left(\frac{(e^{-Q_j\eta}z_j)^{\frac{1}{2}}}{1-e^{-Q_j\eta}z_j}\right)^{d_j-g+1} \exp[Q_j^2 \theta \left(\frac{1}{2} + \frac{e^{-Q_j\eta}z_j}{1-e^{-Q_j\eta}z_j}\right)] \,.
\label{eq:A-roof-over-e-vortex}
\eea

\subsubsection{Chern-Simons Terms}

The supersymmetric Chern-Simons terms generate holomorphic line bundles on the moduli space $\mathfrak{M}_{\fm,i} \cong \Sigma_{d_i}$ according to the general mechanism in~\cite{Collie:2008mx}. A careful translation into the algebraic framework of this paper leads to the conclusion that the Chern-Simons levels $k$, $k_R$ generate holomorphic line bundles $\cL^k$, $\cL_R^{k_R}$ with
\bea
c_1(\cL) & = \theta-\mathfrak{m} \eta \, , \\
c_1(\cL_R) & = -(g-1)\eta \, .
\eea
The contribution to the integrand of equation~\eqref{eq:vortex-general-form} is therefore
\be
\ch(\cL^k \otimes \cL_R^{k_R}) = e^{k (\theta - \mathfrak{m}\eta)} e^{-k_R(g-1)\eta} \, .
\ee
This result passes a consistency check. It is compatible with the contribution~\eqref{eq:A-roof-over-e-vortex} from massive fluctuations of chiral multiplets and the fact that integrating out a massive chiral multiplet of charge $Q_j$ and R-charge $r_j$ with real mass $m_j \to \pm\infty$ shifts
\bea
k & \to k \pm \frac{Q_j^2}{2} \, ,\\
k_{R} & \to k_{R} \pm \frac{Q_j}{2}(r_j-1) \, .
\label{eq:shifts}
\eea

\subsection{Evaluation of Witten Index}

Collecting the the contributions from the tangent directions to the moduli space, the fluctuations of massive chiral multiplets and the supersymmetric Chern-Simons terms, the contribution~\eqref{eq:vortex-general-form} to the twisted index from the component $\cM_{\mathfrak{m},i}$ of the vortex moduli space is 
\begin{align}
	\mathcal{I}_{\fm,i} =\int_{\Sigma_{d_i}}  e^{k (\theta - \mathfrak{m}\eta)} e^{-k_R(g-1)\eta}   &\left(\frac{\eta e^{-\eta/2}}{1-e^{-\eta}}\right)^{d_i-g+1}    \exp[\theta\left(\frac{1}{2}-\frac{1}{\eta}+\frac{e^{-\eta}}{1 - e^{-\eta}}\right)]    \times \nonumber \\
	\prod_{j\neq i}^N & \left(\frac{(e^{-Q_j\eta}z_j)^{\frac{1}{2}}}{1-e^{-Q_j\eta}z_j}\right)^{d_j-g+1}   \exp[ Q_j^2 \theta \left(\frac{1}{2} + \frac{e^{-Q_j\eta}z_j}{1-e^{-Q_j\eta}z_j}\right)]    \, ,
\label{eq:vortex-integral-before-conversion}
\end{align}
provided $\tau' > \mathfrak{m}$, and vanishes otherwise.

The final step is to convert the integration over the symmetric product into a contour integral using the following useful result~\cite{thaddeus1994stable},
\begin{equation}
	\int_{\Sigma_d} A(\eta) e^{\theta B(\eta)} = \oint_{u=0} \frac{\dd u}{u} \, \frac{A(u) \, [1+u \, B(u)]^g}{u^d} \,.
\end{equation}
The integral in equation~\eqref{eq:vortex-integral-before-conversion} has precisely this form with
\begin{align}
	A(\eta) &= e^{-k\mathfrak{m}\eta} e^{-k_R(g-1)\eta} \left(\frac{\eta e^{-\eta/2}}{1-e^{-\eta}}\right)^{d_i-g+1} \prod_{j \neq i}   \left[\frac{(e^{-Q_j\eta}z_j)^{\frac{1}{2}}}{1-e^{-Q_j\eta}z_j}\right]^{d_j-g+1}  \,,\\
	B(\eta) &= k+ \left(\frac{1}{2}-\frac{1}{\eta}+\frac{e^{-\eta}}{1 - e^{-\eta}}\right) + \sum_{j\neq i}  Q_j^2\left(\frac{1}{2} + \frac{e^{-Q_j\eta}z_j}{1-e^{-Q_j\eta}z_j}\right)  \, , 
\end{align}
and therefore we find
\begin{align}
	\cI_{\fm,i} & =\oint_{u=0} \dd u \,  e^{-k\mathfrak{m}u} e^{-k_R(g-1)u} \, \left(\frac{e^{-u/2}}{1-e^{-u}}\right)^{d_i-g+1}   \prod_{j\neq i}^N  \left[\frac{(e^{-Q_ju}z_j)^{\frac{1}{2}}}{1-e^{-Q_ju}z_j}\right]^{d_j-g+1}   \times \nonumber \\
	& \hspace{50pt} \left[k+ \frac{1}{2}\left(\frac{1+e^{-u}}{1 - e^{-u}}\right) + \sum_{j\neq i}^N  \frac{Q_j^2}{2}\left(\frac{1 + e^{-Q_ju}z_j}{1-e^{-Q_ju}z_j}\right) \right]^g \\
	& =\oint_{x=y_i^{-1}} \frac{\dd x}{x} \, x^{k\mathfrak{m}+k_R(g-1)} \, \left(\frac{(x y_i)^{\frac{1}{2}}}{1-x y_i}\right)^{d_i-g+1}   \prod_{j\neq i}^N    \left[\frac{(x^{Q_j}y_j)^{\frac{1}{2}}}{1-x^{Q_j}y_j}\right]^{d_j-g+1}    \times \nonumber \\
	&  \hspace{50pt} \left[k+\frac{1}{2}\left(\frac{1+x y_i}{1 - x y_i}\right) + \sum_{j\neq i}^N  \frac{Q_j^2}{2}\left(\frac{1+x^{Q_j}y_j}{1-x^{Q_j}y_j}\right) \right]^g \,, 
\label{eq:vortex-contribution-for-1}
\end{align}
where the substitution $e^{-u} = x y_i$ has been made in the second line.
A similar calculation can be performed in the case $Q_i = -1$. 

The final result, under the assumption that $|Q_i|=1$ is that the contribution to the twisted index from vortex saddle points parameterised by $\cM_{\fm,i}$ is
\begin{align}
	\mathcal{I}_{\fm,i}=\oint_{x=y_i^{-1/Q_i}} \frac{\dd x}{x} \, x^{k\mathfrak{m}+k_R(g-1)}  \, \prod_{j=1}^N \left[\frac{(x^{Q_i}y_j)^{\frac{1}{2}}}{1-x^{Q_j}y_j}\right]^{d_j-g+1}       \left[k+\sum_{j=1}^N Q_j^2 \left(\frac{1}{2} + \frac{x^{Q_j}y_j}{1-x^{Q_j}y_j}\right) \right]^g 
\label{eq:vortex-contribution}
\end{align}
when $\sign(\tau'-\fm) = \sign(Q_i)$, and zero otherwise. This  exactly reproduces the contribution to the twisted index from the pole at $x^{Q_i}y_i = 1$ when the auxiliary parameter $\eta$ is chosen such that $\sign(\eta)  = \sign(\tau' - \fm)$.

\section{Topological Saddles}
\label{sec:topological}

\subsection{Moduli Space}

Topological saddle points are configurations where $\phi_j = 0$ for all $j=1,\ldots,N$ and there is a unique finite expectation value for $\sigma_0 : = t^2\sigma$ that solves the equation
\begin{equation}
		\tau' - \mathfrak{m} = - \frac{e^2\vol(\Sigma)}{4\pi^2} \sigma_0 k_{\eff}^\pm  \,
	\end{equation}
in the region $\pm \sigma_0 >0$. Topological saddle points exist provided $k_{\text{eff}}^\pm \neq 0$ and $\sign(\tau'-\fm) = \sign(Q_\pm)$. If we choose the auxiliary parameter such that $\sign(\eta) = \sign(\tau'-\fm)$, there are topological saddle points with $\pm \sigma_0 > 0$ whenever the Jeffrey-Kirwan residue prescription selects the poles at $x^{\pm1} \to 0$. The task in this section is to reproduce the residues at these poles.

The only massless bosonic fluctuations around a topological saddle are those of the gauge connection $A$. Topological saddle points with flux $\mathfrak{m}\in\mathbb{Z}$ are therefore  parametrised by connections $A$ on a principle $U(1)$ bundle satisfying
\begin{equation}
*F_A = \frac{2\pi }{\vol(\Sigma)} \mathfrak{m} \,,
\label{eq:top-sol}
\end{equation}
modulo gauge transformations on $\Sigma$. As for vortex saddle points, the contribution to the twisted index is expected to be the Witten index of a supersymmetric quantum mechanics whose target is the moduli space of solutions to these equations. However, gauge transformations act trivially on $F_A$ and $\sigma_0$, so the $U(1)$ gauge symmetry is unbroken and the  quantum mechanics is gauged.

To describe the supersymmetric quantum mechanics concretely, we use the algebraic description of solutions to $\eqref{eq:top-sol}$ as holomorphic line bundle $L$ of degree $c_1(L) = \mathfrak{m}$. We then expect a supersymmetric sigma model to the Picard variety $\text{Pic}^\fm(\Sigma)$, parametrised by the complex structure $\bar\partial_A$ which transforms as a chiral multiplet under $\cN=(0,2)$ supersymmetry. 

However, any holomorphic line bundle has a $\mathbb{C}^*$ worth of automorphisms, corresponding to unbroken complexified gauge transformations. It is therefore more appropriate to describe the supersymmetric quantum mechanics as a sigma model to the Picard stack,
\be
\mathfrak{M}_{\mathfrak{m}} = \mathfrak{Pic}^{\mathfrak{m}}(\Sigma) \, .
\ee
We can make this more concrete at the cost of introducing an auxiliary base point $p \in \Sigma$. Decomposing complex gauge transformations into those trivial at $p$ and constant gauge transformations, we have
\begin{equation}
	\mathfrak{M}_\mathfrak{m}  = \cM_\fm \times [\text{pt} / \mathbb{C}^*] \, ,
\end{equation}
where 
\be
\cM_\fm = \pic{\mathfrak{m}}{\Sigma} \simeq T^{2g}\, .
\ee
In this way, the supersymmetric quantum mechanics is a hybrid of a non-linear sigma model with target space $T^{2g}$ and a $U(1)$ gauge theory.

The supersymmetric quantum mechanics is not, however, a product due to the massive fluctuations of the chiral multiplets $\Phi_j$. They transform in a perfect complex on $\mathfrak{M}_\fm$ generated by fluctuations annihilated by $\bar\partial_A$. Choosing an auxiliary base point as above, this becomes a $\mathbb{C}^*$-equivariant complex on $\cM_\fm$. So the fluctuations roughly transform as sections of a holomorphic vector bundle on the target space $T^{2g}$ of the sigma model and are also charged under the unbroken $U(1)$ gauge symmetry.

\subsection{Contributions to Index}
\label{sec:top-contributions}

The contributions to the twisted index from topological saddle points can be expressed in the same form as vortex saddle points,
\bea
 \int \hat{A} ( \mathfrak{M}_\fm ) \frac{\text{ch}(\cL^k \otimes \cL_R^{k_R})}{\text{ch}(\hat \wedge^\bullet \mathcal{E}) }  \, ,
 \label{eq:top-contribution}
\eea
where $\mathcal{E}$ is a perfect complex arising from fluctuations of the massive chiral multiplet, and $\cL$, $\cL_R$ are holomorphic line bundles arising from the gauge and mixed gauge-R symmetry Chern-Simons terms.

To make this more precise, we choose an auxiliary base point on $\Sigma$ and decompose the moduli stack $\mathfrak{M}_\fm = \cM_\fm\times [\text{pt} / \mathbb{C}^*]$. The characteristic classes in equation~\eqref{eq:top-contribution} are then to be understood as $\mathbb{C}^*$-equivariant classes on $\cM_\fm$. The integral over the moduli stack decomposes into two parts:
\begin{itemize}
	\item A regular integral over the moduli space $\cM_\fm \cong \pic{\mathfrak{m}}{\Sigma}$. This is the usual contribution from an $\cN=(0,2)$ supersymmetric non-linear sigma model.
	\item A contour integral 
	$$
	\frac{1}{2\pi i} \oint_{\mathcal{C}} \frac{\text{d}x}{x} \, ,
	$$
	where $x$ is the Chern character of the trivial $\C^*$-equivariant holomorphic vector bundle with weight $+1$. This is the contribution due to the unbroken $U(1)$ gauge symmetry.
\end{itemize}
The purpose of the contour integral is of course to project onto gauge invariant contributions. This is not meaningful as it stands because the integrals of $\mathbb{C}^*$-equivariant classes in equation~\eqref{eq:top-contribution} over the moduli space $\cM_\fm$ produce rational functions of $x$. It is therefore necessary to specify whether the integrand should be expanded inside or outside the unit circle, which correspond to the residues at $x = 0$ or $x = \infty$ respectively. 

Our prescription will be guided by by physical intuition. First, note that the path integral construction identifies $x = e^{-2\pi \beta (\sigma+iA)}$ where $\sigma$ is the real vectormultiplet scalar and $A$ is a constant gauge connection around the circle. Topological saddle points with $\sigma_0>0$ are therefore associated with the region $x \to 0$, while those with $\sigma_0 <0$ are associated with $x\to \infty$. The natural expectation for the contour $\cC$ is therefore 
\bea
& \sigma_0 > 0: \quad \frac{1}{2\pi i}\int_{x = 0} \frac{\text{d}x}{x} \,, \\
& \sigma_0 < 0: \quad \frac{1}{2\pi i}\int_{x=\infty}
\frac{\text{d}x}{x} \, .
\label{eq:top-res}
\eea

This gains further support from the hamiltonian interpretation of the twisted index as counting supersymmetric ground states. The supersymmetric ground states depend on the sign of the real mass of fluctuations, which is dominated by $\sigma$ as $|\sigma| \to \infty$. For example, the ground state wavefunctions of a 1d chiral multiplet of charge $+1$ are
\bea
& \sigma > 0 : \quad \phi^n e^{-\sigma |\phi|^2}\, , \quad n \geq 0 \,, \\ 
& \sigma < 0 : \quad \bar\phi^n e^{-\sigma |\phi|^2} \bar\psi \, ,\quad n \geq 0
\eea
with contributions to the index
\bea
 \sigma > 0 & : \quad 1 + x + x^2 +\cdots  = \frac{1}{1-x} \,, \\
 \sigma < 0 & : \quad -x^{-1}-x^{-2} + \cdots = \frac{1}{1-x} \, .
\eea
So projecting onto uncharged states at the level of the index is equivalent to
\bea
\sigma>0 & : \quad \frac{1}{2\pi i}\int_{x = 0} \frac{\text{d}x}{x} \frac{1}{1-x}  = 1 \,, \\
\sigma< 0 & : \quad  \frac{1}{2\pi i}\int_{x=\infty} \frac{\text{d}x}{x} \frac{1}{1-x} = 0 \,,
\label{eq:res-example}
\eea
which select the coefficient of $x^0$ in the expansions around $x = 0$ and $x =\infty$ respectively. The general prescription~\eqref{eq:top-res} is basically a broad generalisation of this observation.

In summary, we have two contributions from potential topological vacua with $\sigma_0 >0$ and $\sigma_0<0$ are given by the following integrals
\bea
	I_0 & = \frac{1}{2\pi i}\int_{x = 0} \frac{\text{d}x}{x}  \int \hat{A} ( \cM_\fm ) \frac{\text{ch}(\cL^k \otimes \cL_R^{k_R})}{\text{ch}(\hat \wedge^\bullet \mathcal{E}) }  \, ,\\
	I_\infty & = \frac{1}{2\pi i}\int_{x = \infty} \frac{\text{d}x}{x}  \int \hat{A} ( \cM_\fm ) \frac{\text{ch}(\cL^k \otimes \cL_R^{k_R})}{\text{ch}(\hat \wedge^\bullet \mathcal{E}) }  \, ,
\eea
where we interpret $\mathcal{E}$ and $\cL$, $\cL_R$ as $\mathbb{C}^*$-equivariant objects on the moduli space $\cM_\fm \cong T^{2g}$. In the next section we evaluate these explicitly and show that they reproduce the appropriate contributions to the twisted index according to the contour prescription~\eqref{eq:JK-charges-boundary}.

\subsubsection{Tangent Directions}

Let us first summarise some notation for intersection theory on the Picard variety $\cM_\fm \simeq T^{2g}$. The cohomology ring is generated by classes $\zeta_a \in H^{1,0}(T^{2g},\mathbb{Z})$ and $\bar\zeta_a \in H^{0,1}(T^{2g},\mathbb{Z})$ with $a = 1,\ldots g$. We define $\theta_a := \zeta_a \wedge \bar\zeta_a$ and $\theta := \sum_{a=1}^g \theta_a$ with normalisation
\be
\int_{T^{2g}}\frac{\theta^g}{g!} = 1 \, .
\ee
The tangent bundle is flat and therefore $\hat A(\cM_\fm)=1$.

\subsubsection{Index Bundle}

We now consider the massive fluctuations arising from each of the chiral multiplets $\Phi_j$. At a point on the moduli space corresponding to a holomorphic line bundle $L$, each chiral multiplet generates chiral and Fermi multiplet fluctuations solving 
\bea
\bar\partial_A \phi_j = 0  \, , \quad
\bar\partial_A \eta_j = 0 \, ,
\eea
where $\phi_j$ and $\eta_j$ transform as $0$-form and $1$-form sections of $L^{Q_j} \otimes K_\Sigma^{r_j/2}$ respectively. The  fluctuations of $\Phi_j$ around this point therefore generate the vector spaces
\begin{itemize}
\item Chiral multiplets: \quad $E_j^0 := H^0\left(\Sigma,  L^{Q_j} \otimes K_\Sigma^{r_j/2}\right)\,,$
\item Fermi multiplets: \quad $E_j^1 : = H^1\left(\Sigma,L^{Q_j} \otimes K_\Sigma^{r_j/2}\right)\,.$
\end{itemize}
As $L$ varies in $\text{Pic}^\fm(\Sigma)$ the dimensions of these vector spaces may jump, but by the Riemann-Roch theorem the difference is constant and equal to
\begin{align}
	\dim E_j^0 - \dim E_j^1 
	&= Q_j \mathfrak{m}+ (g-1)(r_j-1) \\
	& = d_j -g +1 \,.
\label{eq:dim-V_i}
\end{align}
This means the difference behave like the fibre of a holomorphic vector bundle on $\text{Pic}^{\fm}(\Sigma)$ for the purpose of K-theoretic computations involved in the twisted index. 

To make this more precise, it is again useful to consider a universal construction. This is canonical for the moduli stack but for concreteness we pick a base point $p \in \Sigma$ and pass the moduli space $\cM_\fm$. There is a corresponding diagram
\begin{equation}
\begin{tikzcd}[column sep=small]
	& { \cM_\fm\times \Sigma  } \arrow[dl,"\pi", swap]\arrow[dr,"p"] & \\ {\cM_\fm} && {\Sigma}
\end{tikzcd} 
\label{eq:pic-projections}
\end{equation}
and universal line bundle $\mathcal{L}$ such that on restriction to a point on $\cM_\fm$ corresponding to a holomorphic line bundle $L$, $\mathcal{L}|_p \simeq L$. The universal line bundle is not unique: due to $\mathbb{C}^*$ automorphisms there is the possibility to transform $\mathcal{L} \to \mathcal{L} \otimes \pi^*\mathcal{N}$. However, this can be fixed by demanding $\cL$ is trivial on restriction to $p$.~\footnote{There is a unique universal line bundle on $\mathfrak{M}_\fm \times \Sigma$ without such a choice.} We also define $\mathcal{K} = p^*K_\Sigma$.

The massive fluctuations of the chiral multiplet $\Phi_i$ generate a perfect complex $\mathcal{E}^\bullet_i$ of sheaves defined by the derived push-forward 
\begin{equation}
	 \mathcal{E}^\bullet_j   := R^\bullet \pi_*\left(\mathcal{L}^{Q_j} \otimes \mathcal{K}^{r_j/2}\right) \,.
\label{eq:def-V_i-topological}
\end{equation}
The stalks of $\mathcal{E}^\bullet_j$ at $L \in \cM_\fm$ are the vector spaces $E_j^\bullet$ considered above. The class $ \ch\left(\mathcal{E}^\bullet_j\right)= \ch(\mathcal{E}^0_j) - \ch(\mathcal{E}^1_j)$ makes sense in equivariant K-theory and the complex behaves like a vector bundle of rank $d_j-g+1$ for the purpose of such computations.

To compute the contribution to the twisted index, we begin by computing the Chern character of $\mathcal{E}^\bullet_j$. This is a small modification of a standard argument presented in~\cite{arbarello1985geometry}. In what follows, we again abuse notation and identify the class $\theta$ with its pull-back via $\pi$. Similarly $\eta_{\scriptscriptstyle \Sigma}$ denotes the class of a point on $\Sigma$ and its pull-back via $p$.

First, the Chern class of the universal line bundle is
\be
c_1(\cL) = \mathfrak{m} \eta_\Sigma + \gamma \, ,
\ee
where $\gamma^2 = -2 \eta_{\scriptscriptstyle \Sigma} \theta$. We therefore find
\begin{align}
	\ch\left(\mathcal{L}^{Q_i}\right)  &= e^{Q_i \mathrm{c}_1(\mathcal{L})} \nonumber\\
	&= 1 + Q_i \mathfrak{m}\eta_{\scriptscriptstyle \Sigma} + Q_i \gamma + \frac{Q_i^2}{2} \gamma^2 \nonumber \\
	&= 1 + Q_i \mathfrak{m}\eta_{\scriptscriptstyle \Sigma} + Q_i \gamma - Q_i^2 \eta_{\scriptscriptstyle \Sigma} \theta \,.
\label{eq:ch-L-Q}
\end{align}
Similarly, from
\begin{equation}
	\mathrm{c}_1(\cK) = (2g-2) \eta_\Sigma\,,
\end{equation}
we find
\begin{align}
	\ch\left(\mathcal{K}^{r_j/2}\right) 
	&= e^{\frac{r_j}{2} \mathrm{c}_1(\mathcal{K})} \nonumber \\
	&= e^{r_j (g-1) \eta_{\scriptscriptstyle \Sigma}} \nonumber \\
	&= 1 + r_j(g-1) \eta_{\scriptscriptstyle \Sigma} \,.
\label{eq:ch-kappa-r-over-2}
\end{align}
Combining these results
\begin{equation}
	\ch\left(\mathcal{L}^{Q_j} \otimes \mathcal{K}^{\frac{r_j}{2}}\right) 
	= 1 + d_j\eta_{\scriptscriptstyle \Sigma} + Q_j \gamma - Q_j^2 \eta_{\scriptscriptstyle \Sigma} \theta \,.
\end{equation}

We can now compute the Chern character of $\mathcal{E}^\bullet_j$ using the Groethendiek-Riemann-Roch theorem,
\begin{align}
	\ch\left(\mathcal{E}^\bullet_j\right) 
	& = \pi_* \left[ \text{ch}(\cL^{Q_j} \otimes \mathcal{K}^{\frac{r_j}{2}} ) \, \text{Td} (\text{Pic}^\fm\Sigma \times \Sigma) \right] \\
	&= \pi_*\left[  \,   \ch\left(\mathcal{L}^{Q_i} \otimes \mathcal{K}^{\frac{r_i}{2}}\right)(1-(g-1)\eta_{\scriptscriptstyle \Sigma}) \right] \nonumber\\
	&= \pi_*\left[1+(d_j-g+1)\eta_{\scriptscriptstyle \Sigma} + Q_i \gamma - Q_i^2 \eta_{\scriptscriptstyle \Sigma} \theta\right] \nonumber \\
	&= (d_j-g+1) - Q_i^2 \theta \\
	& = (d_j-2g+1) + \sum_{a=1}^g e^{-Q_j^2\theta_a} \, ,
\label{eq:Chern-V_i}
\end{align}
where in the final line we have expressed the result in such a way that the Chern roots are manifest.
This is promoted to an equivariant Chern character
\be
\ch\left(\mathcal{E}^\bullet_j\right) = x^{Q_j}y_j \left((d_j-2g+1) + \sum_{a=1}^g e^{-Q_j^2\theta_a} \right) \, .
\ee

The contribution to the twisted index is now given by the equivariant $\hat A$-genus of the complex $\mathcal{E}_j^\bullet$. This is straightforward to compute from the equivariant Chern roots by a now familiar set of manipulations,
\begin{align}
\frac{\hat{A}(\mathcal{E}_j^\bullet)}{e(\mathcal{E}_j^\bullet)} \nonumber 
& = \left(x^{Q_j}y_j\right)^{\frac{d_j-g+1}{2}} e^{\frac{Q_j^2 \theta}{2}}     \left( \frac{1}{1-x^{Q_j}y_j}  \right)^{d_j-2g+1}
\prod_{a=1}^{g} \frac{1}{1-x^{Q_j}y_je^{Q_j^2 \theta_a}}    \nonumber \\
& = \left(x^{Q_j}y_j\right)^{\frac{d_j-g+1}{2}} e^{\frac{Q_j^2 \theta}{2}}     \left( \frac{1}{1-x^{Q_j}y_j}  \right)^{d_j-2g+1}
\prod_{a=1}^{g} \frac{1}{1-x^{Q_j}y_j(1 + Q_j^2 \theta_a)}    \nonumber \\
& = \left(x^{Q_j}y_j\right)^{\frac{d_j-g+1}{2}} e^{\frac{Q_j^2 \theta}{2}}     \left( \frac{1}{1-x^{Q_j}y_j}  \right)^{d_j-g+1}
\prod_{a=1}^{g} \left(1+\frac{x^{Q_j}y_j}{1-x^{Q_j}y_j} Q_j^2 \theta_a)  \right) \nonumber \\
& = \left(x^{Q_j}y_j\right)^{\frac{d_j-g+1}{2}} e^{\frac{Q_j^2 \theta}{2}}     \left( \frac{1}{1-x^{Q_j}y_j}  \right)^{d_j-g+1}
\exp \left( \frac{x^{Q_j}y_j}{1-x^{Q_j}y_j} Q_j^2 \theta   \right) \nonumber \\
&    =  \left( \frac{(x^{Q_j}y_j)^{\frac12}}{1-x^{Q_j}y_j}  \right)^{d_j-g+1}
\exp \left( \left( \frac{1}{2}+\frac{x^{Q_j}y_j}{1-x^{Q_j}y_j} \right)Q_j^2 \theta   \right) \, ,
\label{eq:A-roof-over-e}
\end{align}
where we have made repeated use of $\theta^2_a = 0$.

\subsubsection{Chern-Simons Term}

The supersymmetric Chern-Simons again induce holomorphic line bundles over the moduli space $\cM_\fm \cong T^{2g}$. In the algebraic framework the Chern-Simons levels $k$, $k_R$ induce holomorphic line bundle $\cL^k$, $\cL_{R}^{k_R}$ with
\bea
c_1(\cL) & = \theta \, ,\\
c_1(\cL_R) & = 0 \, ,
\eea
and transform equivariantly with weights $\fm$ and $(g-1)$ respectively. The equivariant Chern characters are therefore
\bea
\text{ch}(\cL^k) & = (x^\fm e^\theta)^k = x^{k\fm}e^\theta \,, \\
\text{ch}(\cL_R^{k_R}) & = x^{k_R(g-1)} \, .
\eea
This is compatible with the contribution~\eqref{eq:A-roof-over-e} from fluctuations of $\Phi_j$ and the fact that integrating out a massive chiral multiplet of charge $Q_j$ and R-charge $R_j$ shifts the supersymmetric Chern-Simons levels as in equation~\eqref{eq:shifts}.

\subsection{Evaluation of Witten Index}

Combining all these contributions, the contribution to the integrand from the integration over the moduli space $\cM_\fm = \text{Pic}^\fm(\Sigma) \cong T^{2g}$ is
\bea
\, & \int_{\text{Pic}^\fm(\Sigma)} x^{k\mathfrak{m}}e^{k\theta} x^{k_R(g-1)} \prod_{j=1}^N \left( \frac{(x^{Q_j}y_j)^{\frac12}}{1-x^{Q_j}y_j}  \right)^{d_j-g+1}
\exp \left( \left( \frac{1}{2}+\frac{x^{Q_j}y_j}{1-x^{Q_j}y_j} \right)Q_j^2 \theta   \right) \\
& = x^{k\fm+k_R(g-1)} \prod_{j=1}^N \left( \frac{(x^{Q_j}y_j)^{\frac12}}{1-x^{Q_j}y_j}  \right)^{d_j-g+1} \int_{\text{Pic}^\fm(\Sigma)}  
\exp \left(k+\sum_{j=1}^N Q_j^2\left( \frac{1}{2}+\frac{x^{Q_j}y_j}{1-x^{Q_j}y_j} \right)    \right)\theta \\
& = x^{k\fm+k_R(g-1)} \prod_{j=1}^N \left( \frac{(x^{Q_j}y_j)^{\frac12}}{1-x^{Q_j}y_j}  \right)^{d_j-g+1}  
 \left[k+\sum_{j=1}^N Q_j^2\left( \frac{1}{2}+\frac{x^{Q_j}y_j}{1-x^{Q_j}y_j} \right)    \right]^g \, .
\eea
The contributions from topological saddle points therefore exactly reproduce the potential residues at $x = 0$ and $x = \infty$ in the Jeffrey-Kirwan residue prescription with $\eta$ aligned with $\tau'-\fm$.

\section{Examples}
\label{sec:examples}

We consider a $U(1)$ Chern-Simons theory at level $k \in \frac{1}{2}+\mathbb{Z}_{\geq0}$ with one chiral multiplet $\Phi$ of R-charge $r=1$ and charge $Q=+1$. The flavour symmetry $T_\text{f}$ is trivial and there are no real mass parameters. The effective Chern-Simons level coupling is
\be
k_\text{eff}(\sigma) = k + \frac{1}{2} \, \sign( \sigma) 
\ee
and so
\be
k_\text{eff}^\pm = k \pm\frac{1}{2} \, .
\ee
The cases $k=\frac{1}{2}$ and $k > \frac{1}{2}$ are quite different. The former has a neutral monopole operator and is mirror to a free chiral multiplet. This difference is reflected in the structure of the saddle points in our computation of the twisted index and therefore we treat the two cases separately. We also restrict attention to the twisted index with $g>0$.

\subsection{$U(1)_{\frac{1}{2}}$ + $1$ Chiral}
\label{subsec:Example-U(1)-1/2-1-chiral}

First consider $k = \frac{1}{2}$. In this case $k_\text{eff}(\sigma) = \frac{1}{2}(1+\sign( \sigma) )$ and therefore $k_\text{eff}^+=1$ and $k_\text{eff}^- = 0$. There is a neutral monopole operator and the theory is mirror to a free chiral multiplet, together with specific background mixed Chern-Simons couplings.

The contour integral for the twisted index is
\begin{equation}
	\mathcal{I} = \sum_{\mathfrak{m}\in \mathbb{Z}}  \frac{(-q)^\mathfrak{m}}{2 \pi i} \oint_\mathcal{C} \frac{\dd x}{x} \, \frac{x^{\fm}}{(1-x)^{\fm+g}}\,.
\label{eq:U(1)-k-1/2-index}
\end{equation}
where we have shifted $q\to-q$ compared to above. In the presence of a 1d Fayet-Iliopoulos parameter $\tau$, the contour is a Jeffrey-Kirwan residue prescription with charges
\be
Q_{+} = -1 \, , \quad Q_1 = 1 \, ,\quad Q_- = \fm-\tau' \, .
\label{eq:example-charges-1}
\ee
Note that the charge $Q_-$ associated to the residue at $x=\infty$ now depends on the 1d Fayet-Iliopoulos parameter $\tau$ according to equation~\eqref{eq:JK-charges-boundary} since $k_\text{eff}^-=0$.

For $g>0$ the residue at $x = \infty$ vanishes and there is no wall-crossing phenomena. The twisted index is given by computing the residue at $x=1$ ($\eta>0)$ or equivalently minus the residue at $x=0$ ($\eta<0)$, with the result
\be
\cI = (-1)^g q^{1-g}(1-q)^{g-1}\, .
\ee
While the twisted index is non-zero only for fluxes $1-g\leq \mathfrak{m}\leq 0$, there are in fact supersymmetric ground states for all $\fm \geq 1-g$~\cite{Bullimore:2018yyb}. 

We now reproduce this result by evaluating the contributions from vortex and topological saddle points. The existence of vortex and topological saddle points is constrained by equation \eqref{eq:integrated-vortex-equation}, which becomes
\begin{equation}
\left(\tau' - \mathfrak{m}\right) +\frac{e^2\vol(\Sigma)}{4\pi^2}\sigma_0 k_{\eff}(\sigma) =  \norm{\phi}^2 \,,
\end{equation}
together with the equation $\sigma \phi = 0$. The existence of solutions depends on the sign of $\tau'-\fm$.
\begin{itemize}
\item When $\tau'-\fm>0$, there are vortex saddle points with $\sigma_0 = 0$. The moduli space of vortex solutions with flux $\mathfrak{m}$ is the symmetric product $\mathfrak{M}_\fm = \Sigma_d$ where $d = \fm+g-1$. Following the computations in section~\ref{sec:vortex}, the contribution to the twisted index is
\bea
\int_{\Sigma_d} \hat{\text{A}}(T\Sigma_d) \, \ch(\cL^{1/2})
& = \int_{\Sigma_d} \, \left(\frac{\eta e^{-\eta}}{1-e^{-\eta}}\right)^{\fm}    \exp[\theta\left(1-\frac{1}{\eta}+\frac{e^{-\eta}}{1 - e^{-\eta}}\right)] \\
& = \frac{1}{2\pi i} \int_{x = 1} \frac{\dd x}{x} \frac{x^\fm}{(1-x)^{\fm+g}}\, .
\eea
\item When $\tau'-\fm<0$, there are topological saddle points with
\be
\phi=0\, , \quad \sigma_0 = - \frac{4\pi^2}{e^2\text{vol}(\Sigma)}(\tau' - \mathfrak{m}) > 0 \, .
\ee
The moduli space of topological solutions with flux $\mathfrak{m}$ is the Picard variety $\mathfrak{M}_\fm = \text{Pic}^{\fm}\Sigma$. Following the computations in section~\ref{sec:topological}, the contribution to the twisted index is
\bea
\int_{\Sigma_d} \hat{\text{A}}(T\Sigma_d) \, \ch(\cL^{1/2}) 
& = \frac{1}{2\pi i} \int_{x=0} \frac{\dd x}{x} \int_{\text{Pic}^\fm\Sigma} \, \left(\frac{x}{1-x}\right)^{\fm}    \exp[\left(\frac{1}{1 - x}\right)\theta ] \\
& = \frac{1}{2\pi i} \int_{x =0} \frac{\dd x}{x} \frac{x^\fm}{(1-x)^{\fm+g}} \, ,
\eea
where the residue at $x=0$ is taken since $\sigma_0>0$.
\end{itemize}
A Coulomb branch of solutions with $\sigma_0 <0$ opens at $\tau'-\mathfrak{m}=0$ so there is the potential for wall-crossing. However, the vanishing of the residue at $x = \infty$ means that the twisted index is independent of $\tau$. This reproduces the Jeffrey-Kirwan residue prescription with charges~\eqref{eq:example-charges-1} and $\sign(\eta) = \sign( \tau'-\mathfrak{m})$. The result is independent of $\eta$ for each flux $\fm$ by construction.

\subsection{$U(1)_k$ + $1$ Chiral}

Now consider $k>\frac{1}{2}$ such that $k_{\text{eff}}^\pm = k \pm \frac{1}{2} > 0$. There are no gauge neutral monopole operators and the structure of the twisted index differs considerably.

The contour integral for the twisted index is now
\begin{equation}
	\mathcal{I} = \sum_{\mathfrak{m}\in \mathbb{Z}}  \frac{(-q)^\mathfrak{m}}{2 \pi i} \oint_\mathcal{C} \frac{\dd x}{x} \, x^{k\mathfrak{m}} \, \left( \frac{x^{1/2}}{1-x}\right)^\fm \left( k+\frac{1}{2}\frac{1+x}{1-x}\right)^g\,,
\label{eq:U(1)-k-1/2-index}
\end{equation}
where the contour is a Jeffrey-Kirwan residue prescription with charges
\be
Q_+ = -k-\frac{1}{2}<0\, ,  \quad Q_1 = 1\, , \quad Q_- = k - \frac{1}{2} > 0 \, .
\label{eq:example-charges-1}
\ee
The index is now manifestly independent of $\tau$ and there is no wall-crossing. We therefore enumerate the residues at $x=1$ and $x=\infty$ ($\eta >0$) or equivalently minus the residues at $x=0$ $(\eta <0)$. 

In this case it is illuminating to spell out the contributions from individual residues. For example, with $g=2$ and $k = \frac{3}{2}$ we find
\bea
-\cI_0 & =\frac{1}{q} - 4 \, , \\
\cI_1 & =\frac{1}{q} - 3 - q - 2q^2 -5q^3-14q^4- \cdots  = \frac{1-4 q+\sqrt{1-4 q}}{2 q} \, , \\
\cI_\infty & = -1 + q + 2q^2 +5q^3+14q^4+ \cdots =  \frac{1-4 q-\sqrt{1-4 q}}{2 q} \, .
\eea
Notice that the contributions $\cI_1$ and $\cI_\infty$ are not rational function of $q$ and so cannot individually reproduce a reasonable index. In fact they do not count honest supersymmetric ground states but only perturbative ground states. These are subject to instanton corrections that remove pairs of perturbative ground states corresponding to cancelations in the sum $\cI_1 +\cI_\infty = - \cI_0$.

We can reproduce these contributions from an analysis of vortex and topological saddle points. 
The saddle points are again constrained by 
\begin{equation}
\left(\tau' - \mathfrak{m}\right) +\frac{e^2\vol(\Sigma)}{4\pi^2}\sigma_0 k_{\eff}(\sigma) =  \norm{\phi}^2 \,,
\end{equation}
and depend on the sign of $\tau'-\mathfrak{m}$.

\begin{itemize}
	\item When $\tau'-\mathfrak{m}> 0$, there are both vortex saddle points and topological saddle points with $\sigma_0 < 0$. The contributions from these saddle points reproduce the residues at $x = 1$ and $x=\infty$ respectively.
	\item When $\tau'-\mathfrak{m}< 0$, there are topological saddle points with $\sigma_0 >0$, whose contribution reproduces the residue at $x = 0$.
\end{itemize}
There is no Coulomb branch at $\tau'-\mathfrak{m} = 0$ and the twisted index is independent of $\tau$. This reproduces precisely the Jeffrey-Kirwan residue prescription with charges~\eqref{eq:example-charges-1} and $\sign(\eta) = \sign(\tau' - \mathfrak{m})$. The result is independent of $\eta$ for each flux $\fm$ by construction.

\section{An Exploration of $SU(2)$}
\label{sec:su2}

In this section, we briefly explore the extensions of our results to $G=SU(2)$ Chern-Simons matter theories, highlighting some novelties and difficulties compared to $G = U(1)$. For simplicity, we focus on $SU(2)$ Chern-Simons theory at level $k$ coupled to $N$ fundamental chiral multiplets.~\footnote{The quantisation condition requires $k + \frac{N}{2} \in\mathbb{Z}$.
In this section, we will assume $N\in 2\mathbb{Z}$ and $k\in\mathbb{Z}$.}

\subsection{The Twisted Index}

The contour integral formula reads
\begin{equation}
	\mathcal{I} = \sum_{\mathfrak{m}\in \mathbb{Z}} \frac{1}{2 \pi i} \oint_{\Gamma} \frac{\dd x}{x} \, H(x)^g Z(x,\mathfrak{m}) \, ,
\label{eq:partition-function-contour-integral-su2}
\end{equation}
where
\begin{equation}
Z(x,\mathfrak{m}) =\frac{1}{2}~x^{2\mathfrak{m}k}\left(1-x^2\right)^{1-g}\left(1-x^{-2}\right)^{1-g}\prod_{i=1}^N \frac{(1-x^{-1}y_i^{-1})^{\mathfrak{m}-(g-1)(r_i-1)}}{(1-xy_i^{-1})^{\mathfrak{m} +(g-1)(r_i-1)}} \, ,
\label{eq:Z-su2}
\end{equation}
with $\mathfrak{m}$ valued in the co-character lattice of $G=SU(2)$. We also have
\begin{equation}
H = \left( 2k + \sum_{i=1}^{N} \frac{1+ xy_i^{-1}}{2(1-xy_i^{-1})} + \frac{1+x^{-1}y_i^{-1}}{2(1-x^{-1}y_i ^{-1})}\right)\, .
\end{equation}
The factor $(1-x^2)^{1-g}(1-x^{-2})^{1-g}$ in equation~\eqref{eq:Z-su2} originates from the $W$-bosons in breaking the gauge group from $SU(2)$ to $U(1)$. The corresponding poles for $g>1$ lead to subtleties in the derivation of the Jeffrey-Kirwan residue prescription. The standard approach in the literature is to omit these poles from the contour $\Gamma$~\cite{Benini:2016hjo,Closset:2016arn}. For the remaining poles, the same prescription as in \eqref{eq:JK} applies. The poles at $x^{\pm1} \to 0$ are assigned charges according to the effective Chern-Simons levels
\bea\label{effective level cartan}
Q_+ &= -k^+_{\mathrm{eff},U(1)} = - 2k\, ,\\
Q_- &= k^-_{\mathrm{eff},U(1)} = 2k\,,
\eea
and the contour encloses poles whose charges are of the same sign of the auxiliary parameter $\eta$. 
This prescription becomes ambiguous when $k=0$. This appears compatible with the geometric interpretation discussed in the remainder of this section, which may provide some further justification for this prescription, but we are unable to provide a systematic explanation.

\subsection{Geometric Interpretation}

In analogy to the $U(1)$ theories discussed above, we now study the saddle point equation in order to interpret the index geometrically. The supersymmetric saddle points are determined by the equations
\begin{subequations}
\be
\label{su2-bps-1}
\frac{1}{e^2} * F_A + \sum_{i=1}^N \phi_i^\dagger \phi_i - t^2\sigma\frac{k_{\text{eff}}(\sigma)}{2\pi} =0 \,, 
\ee
\be
\label{su2-bps-2}
d_A\sigma=0\,,\quad \bar\partial_A\phi_i=0 \,,
\ee
\be
\label{su2-bps-3}
(\sigma + m_i)\phi_i=0 \,,
\ee
\end{subequations}
where $\sigma = \sigma^a T^a$. For each $i$, the chiral multiplet $\phi_i$ transforms as a section of the vector bundle $K_\Sigma^{r_i/2} \otimes E^{i}$. Suppose that the mass $m_i$ is generic. The $SU(2)$ gauge group is unbroken only at $\sigma=0$, where the effective level $k_{\text{eff}}$ is given by
\be\label{effective su2 level}
k_{\text{eff},SU(2)} =  k + \frac12 \sum_{i}T_2(R_i)\text{sign}(m_i)\, .
\ee
For $\sigma\neq0$, the gauge group is broken to $U(1)$ and the effective levels are given by \eqref{effective level cartan}.

These equations exhibit the same types of solutions as their abelian counterparts: vortex and topological saddles. For general $k$, there is an important subtlety due to the existence of topological saddles with both $\sigma=0$ and $\phi=0$. The $SU(2)$ gauge symmetry is unbroken and the relevant moduli space is that of $SU(2)$ flat connections on $\Sigma$. They are not topologically disjoint from vortex and topological saddles where the gauge group is broken to $U(1)$. 

In order to circumvent this issue, in the discussion below, we deform the moduli space by turning on a small 1d Fayet-Iliopoulos parameter $\tau\in\mathbb{R}$ that explicitly breaks $SU(2)$ to $U(1)$. This modifies \eqref{su2-bps-1} to
\be
\frac{1}{e^2} * F_A + \sum_{i=1}^N \phi_i^\dagger \phi_i - t^2\sigma\frac{k_{\text{eff}}(\sigma)}{2\pi} =\tau \mathbf{1}\, .
\ee
For convenience, we take the normalised $\tau$ defined in \eqref{eq:normalised-1d-FI} to be \be|\tau' |<1 \, ,\label{eq:tau-reg}\ee although in principle any value of $\tau'$ is allowed.

For $k\neq 0$, unless $\tau'\notin \mathbb{Z}$, the topological saddles with $SU(2)$ unbroken are removed and the twisted index is well-defined and independent of $\tau$. For $k=0$, there exist a non-compact Coulomb branch parametrised by constant $\sigma_0:=t^2\sigma$ and therefore the twisted index may jump as we vary $\tau'$ across integer values.

\subsubsection{Vortex Saddles}

Vortex saddles are solutions with $\phi_i\neq 0$ for some $i$. Equation \eqref{su2-bps-3} implies that $\sigma^1=\sigma^2=0$ and $\sigma^3 = \pm m_i$. This equation breaks the gauge group $SU(2)\rightarrow U(1)$ by itself. Accordingly, the vector bundle decomposes into a sum of line bundles, whose $i$-th summand is
\be\label{decompose}
E_i = L_i \oplus L_i^{-1}\,.
\ee
We denote by $\phi_{1,i}$ and $\phi_{2,i}$ the sections of $L_i$ and $L_i^{-1}$ respectively. In the limit $t\rightarrow 0$ and in the presence of generic mass parameters $m_i$, the moduli space of vortex saddles is a union of disjoint components $\cM_{{\mathfrak m},i}^+$ and $\cM_{{\mathfrak m},i}^-$, which are parametrised by solutions $(A,\phi)$ that satisfy
\be\label{su2 vortex 1}
\frac{1}{e^2} * F_A + \frac{1}{2} |\phi_{1,i}|^2  =\tau\, , \quad \partial_A \phi_{1,i}=0
\ee
and
\be\label{su2 vortex 2}
\frac{1}{e^2} * F_A - \frac{1}{2}  |\phi_{2,i}|^2  =\tau \,, \quad \partial_A \phi_{2,i}=0
\ee
respectively. Here $A$ is a connection on the line bundle $L_i$ of degree $\frak m$. We then have
\be
 \cM_{\mathfrak{m},i}^+ = \begin{cases}
\Sigma_{d_i^+} & \quad \text{if}\quad  \mathfrak{m}<\tau' \\
\emptyset & \quad \text{if}  \quad \mathfrak{m} > \tau'
\end{cases}\, ,\qquad \cM_{\mathfrak{m},i}^- = \begin{cases}
\Sigma_{d_i^-} & \quad \text{if}\quad  \mathfrak{m}>\tau' \\
\emptyset & \quad \text{if}  \quad \mathfrak{m} < \tau'
\end{cases}\,,
\ee
with 
\be
d_i^\pm : = \pm\mathfrak{m} + r_i(g-1)\, .
\ee

The contribution to the twisted index from the vortex saddles can be studied in the same way as in section \ref{sec:vortex-contributions}. Each component contributes
\be
\cI_i^\pm= \int \hat{A} ( \cM^\pm_{\mathfrak{m},i} ) \frac{ \text{ch}(\cL^{2k})}{\text{ch}(\hat \wedge^\bullet \mathcal{E} ) }\, .
\ee
Here the index bundle $\cE$ is the contribution from the fluctuation of the chiral and vector multiplets:
\be
\cE =  \cE_V^\bullet\otimes \bigotimes_{j\neq i}  \cE_{+,j}^\bullet \otimes  \bigotimes_{j= 1}^N \cE_{-,j}^\bullet \, ,
\ee
where $\cE_{\pm,j}^\bullet$ is the contribution from the chiral multiplets $\phi_{1,i}$ and $\phi_{2,i}$:
\be
\cE_{\pm,j}^\bullet = R^\bullet \pi_*(\mathcal{L}^{\pm} \otimes \mathcal{K}^{r_j /2})\, .
\ee
The first term $\cE_V^\bullet$ is the contribution from fluctuations of W-bosons. This decomposes into 1d $\cN=(0,2)$ chiaral and Fermi multiplets
\begin{itemize}
\item Chiral multiplets $(a_z,\Lambda_z)$: $E_V^1 := H^0(\Sigma,L^2\oplus L^{-2}) \,,$
\item Fermi multiplets $(\sigma,\lambda)$: $E_V^0 := H^1(\Sigma, L^2\oplus L^{-2}) \,,$
\end{itemize}
which can be written in terms of the universal bundle as
\be
\cE_V^\bullet = R^\bullet \pi_*(\cL^2 \oplus \cL^{-2})\, .
\ee
From this we can compute
\bea
\frac{1}{\text{ch}(\hat\wedge^\bullet \cE)} =&~ (e^{-\eta}-e^\eta)^{2\fm-g+1}(e^{\eta}-e^{-\eta})^{-2\fm-g+1}\prod_{j\neq i} \left(\frac{(e^{-\eta}z_j)^{1/2}}{1-e^{-\eta}z_j}\right)^{d_j^+-g+1}\prod_{j=1}^N \left(\frac{(e^{\eta}z_j)^{1/2}}{1-e^{\eta}z_j}\right)^{d_j^--g+1} \\
& ~\text{exp}\left[\theta \sum_{j\neq i} \left(\frac12 + \frac{e^{-\eta} z_j}{1-e^{-\eta}z_j}\right) + \theta \sum_{j=1}^N \left(\frac12 +\frac{e^{\eta} z_j}{1-e^{\eta}z_j}\right)\right]\, ,
\eea
where we defined $z_j = y_j/y_i$ with $y_i = e^{-2\pi  m_i}$ as before.

Finally, the $SU(2)$ Chern-Simons term at level $k$ generates a holomorphic line bundle $\cL_{\fm,i}^{2k}$ on the moduli space $\cM_{\fm,i}^\pm$, which contributes
\be
\text{ch}(\cL_{\fm,i}^{2k}) = e^{2k(\theta -\fm \eta)}\, .
\ee
By the same manipulations that follow~\eqref{eq:vortex-integral-before-conversion}, we get an agreement of the integrand with \eqref{eq:partition-function-contour-integral-su2}. The integration picks the poles of the chiral multiplets according to the alignment of $\eta$ and $\tau'-\fm$, which for $\fm \neq 0$ and $\tau'$ chosen as in \eqref{eq:tau-reg} is the same as $-\fm$. 

\subsubsection{Topological Saddles}

Topological saddles are characterised by $\phi_i=0$ for all $i$. With $\tau$ turned on, we only have solutions with $\sigma \neq 0$ and the vector bundle decomposed into a sum of line bundles $E = L \oplus L^{-1}$.~\footnote{In the limit $\tau \rightarrow 0$, the moduli space of flat $SU(2)$ connections appears, as we discussed above.}

The BPS equations in the limit $t\rightarrow 0$ have solutions of this type in the region $|\sigma|\rightarrow \infty$ with $\sigma_0=t^2\sigma$ kept finite. The equations uniquely determine the value of $\sigma$. Therefore this component of the moduli space is parametrised by connections $A$ of the $U(1)$ bundle $L$ satisfying
\be\label{u1 top}
*F_A = \frac{2\pi}{\text{vol}(\Sigma)}\fm\, ,
\ee
where deg$(L)=\fm$. The low-energy theory is described by the $U(1)$ Chern-Simons theory with
\be
k_{\text{eff},U(1)}^{\pm} = 2k\, .
\ee

The geometric description of the twisted index can be given in the same way as in section \ref{sec:topological}. Algebraically, this topological saddles can be described by the Picard stack \be
\mathfrak{M}_{\mathfrak{m}} = \mathfrak{Pic}^{\mathfrak{m}}(\Sigma) \, , 
\ee
which contributes to the twisted index as
\be
\int \hat A(\mathfrak{M}_{\mathfrak{m}}) \frac{\text{ch}(\Theta^{2k})}{\text{ch}(\hat\wedge^\bullet\cE)}\, .
\ee
As discussed in section \ref{sec:top-contributions}, this integral decomposes into an integral over the moduli space $\cM_\fm$ and a contour integral around $x=0,\infty$ that projects onto $\mathbb{C}^*$-invariant contributions.

The factor $\text{ch}(\hat\wedge^\bullet \cE)^{-1}$ is the contribution from fluctuations of chiral and vector multiplets on ${\frak M}_\fm$. Following the discussion in section \ref{sec:top-contributions}, we find
\bea
\frac{1}{\text{ch}(\hat \wedge^\bullet \cE)} =&~ (x^{-1}-x)^{2\fm -g+1}(x-x^{-1})^{-2\fm -g+1} \prod_{i=1}^N \left(\frac{(x y_i)^{1/2}}{1-xy_i}\right)^{\fm + (g-1)(r_i-1)} \left(\frac{(x^{-1} y_i)^{1/2}}{1-x^{-1}y_i}\right)^{-\fm + (g-1)(r_i-1)}\\
& ~\exp\left[\sum_{i=1}^N \theta \left(\frac12 + \frac{xy_i}{1-x y_i}\right) + \theta \left(\frac12 + \frac{x^{-1}y_i}{1-x^{-1} y_i}\right)\right]\, .
\eea
The effective $U(1)$ CS coupling contributes
\be
\text{ch}(\Theta^{2k})=e^{2k\theta}e^{2k\fm}\, .
\ee
The result agrees once again with \eqref{eq:partition-function-contour-integral-su2}.
Projecting onto $\mathbb{C}^*$-invariant configuration is done by taking residues at $\sigma \rightarrow \pm \infty$ where $x=e^{-2\pi\sigma}$ as usual. Poles are then picked up according to the assignment of charges~\eqref{eq:JK-charges-boundary}, which we rewrite for convenience:
 \begin{subequations}
\begin{align}
	Q_{0} &= 
	\begin{dcases}
    	- 2k \quad &\text{if} \quad k  \neq 0 \\
    	\mathfrak{m}- \tau' \quad &\text{otherwise} 
  	\end{dcases}\,, \\
	Q_{\infty} &= 
	\begin{dcases}
    	+ 2k \quad &\text{if} \quad k \neq 0 \\
    	\mathfrak{m}- \tau' \quad &\text{otherwise} 
  	\end{dcases}\,.
\end{align}
\end{subequations}

The final answer for the index computation may or may not depend on $\tau$, depending on the presence of vectormultiplet poles and on whether the level $k$ vanishes. We conclude by studying some examples on $\mathbb{P}^1$ at level $k\neq 0$, which are immune of these problems.

\subsection{Example}

At genus $g = 0$, all holomorphic vector bundles split as a sum of holomorphic line bundles by the Birkhoff-Groethendieck theorem. Therefore, the topological vacua that preserve full $SU(2)$ gauge group are absent and the subtleties described in the paragraph below \eqref{effective su2 level} do not arise. This is reflected in the fact that there are no vectormulitplet poles in the integrand of the contour formula. The $SU(2)$ theory can then be viewed as a 1d $U(1)$ theory prior to the insertion of the regulator. For $k\neq 0$, we do not expect a dependence of the result on the regulator $\tau$.

Let us consider the example of $G=SU(2)_k$ theory with $N = 2$ fundamental chiral multiplets. The holomorphic vector bundle $E$ decomposes into a sum of line bundle $E = L \oplus L^{-1}$ with deg$(L)=\fm\in\mathbb{Z}$. The D-term equation reduces to
\be
\frac{1}{e^2} * F_A +  \frac{1}{2}\sum_{i=1}^2\left(|\phi_{1,i}|^2 -|\phi_{2,i}|^2\right) -t^2\frac{2k\sigma}{2\pi}   =0\, ,
\ee
We find that the solutions of the BPS equations exists in the following regions, depending on sign of $\fm$\footnote{For convenience, we take the regulator to be small.}:
\begin{itemize}
\item For $\fm>0$, we have vortex solutions at $\sigma=m_i$ and topological solutions at $\sigma\rightarrow \infty$;
\item For $\fm<0$, we have vortex solutions at $\sigma=-m_i$ and topological solutions at $\sigma\rightarrow -\infty$;
\item For $\fm=0$, we have $\left\{\begin{array}{l}
\tau>0: \text{vortex solutions at } \sigma=-m_i \text{ and topological solution at } \sigma \rightarrow -\infty \text{;} \\
 \tau<0: \text{vortex solutions at } \sigma=m_i \text{ and topological solutions at } \sigma\rightarrow \infty \text{.}
\end{array}\right.$
\end{itemize}

On the other hand, the Jeffrey-Kirwan charges for the poles at infinities are assigned as
\be
Q_0 = -2k\,,\quad Q_\infty = 2k \,.
\ee
Let $Q_+$ be the collection of positive $U(1)$ charges at $\sigma=-m_i$, and $Q_-$ be the negative $U(1)$ charges at $\sigma=m_i$. Suppose $\eta>0$. The Jeffrey-Kirwan residue integral then picks up the poles that corresponds to charges $Q_{+}$ and $Q_\infty$ for all values of $\fm\in\mathbb{Z}$ and $\tau$.
By the residue theorem, we find that this prescription agrees with the geometric interpretation of the indices at each flux sector. 

\acknowledgments

MB gratefully acknowledges support from the EPSRC Early Career Fellowship EP/T004746/1 ``Supersymmetric Gauge Theory and Enumerative Geometry".

\bibliographystyle{JHEP}
\bibliography{main}
\end{document}